\def\etal{{\em et al.~}}
\def\H0{$H_0$ = 100 {\it h} km s$^{-1}$ Mpc$^{-1}$}

\documentclass[flushrt]{aastex}
\singlespace
\usepackage{psfig}
\usepackage{natbib}

\slugcomment{Accepted for publication in the {\em Astronomical Journal}}

\shortauthors{Gaspar Galaz}
\shorttitle{E+A galaxies in the near-IR}

\begin{document}

\title{E+A galaxies in the near-IR: Broad band photometry}
\author{Gaspar Galaz\altaffilmark{1}} 
\affil{Carnegie Observatories. Las Campanas 
Observatory, Casilla 601, La Serena, Chile} 
\altaffiltext{1}{Andes-Carnegie Fellow}
\email{gaspar@azul.lco.cl}

\begin{abstract}

We present near-IR photometry of a selected sample of southern hemisphere E+A
galaxies. The sample includes 50 galaxies from 
nearby ($z \sim 0.05$) and distant ($z \sim 0.3$) clusters 
as well as E+A galaxies from the field ($z \sim 0.1$).
We also observed 13 normal early-type galaxies from the field and from clusters
to be compared with the E+A sample. 
The photometry includes $J$, $H$ and $K_s$ apparent magnitudes and colors. 
Observed colors are obtained from the apparent total magnitudes and compared
to spectrophotometric models of galaxy evolution GISSEL96. There is an 
overall agreement between integrated colors of models and observed ones, for 
both the E+A located in clusters and in the field, at $z \lesssim 0.1$. However, 
large differences are found between colors predicted from models and those observed in E+A 
galaxies located in clusters at $z \sim 0.3$. 

We also compute 
rest-frame colors for all the galaxies using two different 
sets of K-corrections, and obtain average colors for all the samples.

We investigate systematic properties of the E+A sample as a 
function of their environment. Results seem 
to indicate that cluster E+As (at low redshift) are bluer than field
E+As at $z \sim 0.1$. Even this conclusion does not depend whether 
we use comoving or rest-frame colors, nor
on the models used to obtain rest-frame colors; the difference is not 
significant enough, considering color dispersions between the samples. 
If differences are real, they could imply different stellar content for the
E+A galaxies located in the field, compared to the cluster E+A. 
\end{abstract}


\section{Introduction}

The attention drawn to E+A galaxies (or post-starbursts) has 
increased since it was claimed that 
their fraction observed in clusters is correlated with the Butcher-Oemler 
effect \citep{butcher78, dressler83, rakos95}.
The E+A galaxies present a peculiar spectrum in the optical:
strong Balmer absorption lines, representative of a large population of A and B
stars, but a {\em lack} of
emission lines typical of blue, star forming galaxies, like 
[OII]$\lambda3727$, [OIII]$\lambda$5007, and H$\alpha$. 
Because of the further detection 
of metallic absorption lines such as Mg b $\lambda 5175$, Ca H \& K 
$\lambda 3934$, 3968 and Fe $\lambda5270$, indicative of an old 
population dominated by G, K and M spectral types they were 
called E+A \citep{dressler83}.
In addition, their spectra in the {\em optical} cannot be reproduced simply by 
a young stellar component without nebular emission lines: it is necessary to add an old
population of stars, like the one present in 
quiescent elliptical galaxies \citep{liu96}. 

During the last 3 years, research on the E+A galaxies has seen a revival
after the discovery of more such galaxies not only in nearby clusters,
like Coma and others (e.g. see \citet{caldwell97} and references
therein), but also in the field; in particular those discovered 
during the Las Campanas Redshift Survey (LCRS, Shectman \etal 1996)
by \citet{zabludoff96}. An interesting feature
is the apparent existence of two classes of E+A galaxies \citep{couch87}. One 
is formed by ``blue'' post-starburst galaxies  
and the other by redder, H$\delta$-strong (HDS)
galaxies \citep{fabricant91}. These subclasses
of E+As have colors and absorption line features related to the 
morphology: HDS E+As have in general a noticeable bulge and/or
spheroidal component when compared with the blue class. In addition, the
HDS class can be divided into 2 subclasses: bulge and disk HDSs (as 
observed in A665 and in Coma \citep{franx93}). 

An almost unexplored domain of these E+A galaxies is their 
near-IR properties. In fact, no catalogue or systematic observations
exist on this subject. Is the bright, red population, dominated mainly 
by giants and stars of the asymptotic giant branch (AGB) of the E+As, different
from the red population of other elliptical galaxies, in particular the 
perturbed ellipticals? Is there any conspicuous signature in the 
near-IR colors of the E+A galaxies, as in the 
optical wavelengths? Do the cluster E+A galaxies have bluer colors
than those in the field at these wavelengths? Are the near-IR colors of the E+A 
galaxies similar to the colors of normal galaxies, as predicted 
by spectrophotometric models? 

In this paper, we investigate these
questions with new data taken at Las Campanas
Observatory, in Chile. The sample includes E+A galaxies from the 
field (most of them from the LCRS) and from nearby clusters as well as 
clusters at $z \sim 0.3$. 

The paper is organized as follows. In \S 2 we present  the 
galaxies selected for this work. In \S 3 we explain the 
observations, and in \S 4 the data reduction procedures. In \S 5 we present
the results of the photometry, the apparent
magnitudes, colors, and K-corrections. Both, {\em observed}  and {\em rest-frame} 
colors are compared
with colors obtained from spectrophotometric models of galaxy evolution in \S 6. In \S 7
we discuss the limitations of our results and the implications for some properties
of the E+A galaxies from our near-IR colors. Conclusions are presented in \S 8. 

\section{The Sample}

All of the galaxies selected for this study have been spectroscopically
classified as E+A galaxies from the analysis of their Balmer absorption lines 
(particularly the equivalent widths of H$\delta$ and H$\beta$) 
and the lack of nebular emission lines, 
representative of an ongoing stellar formation process. We have selected
most of the southern E+A galaxies existent in the literature 
at present. The sample of galaxies
is divided into 4 subsamples. The first subsample corresponds to 21 field 
E+A galaxies from the LCRS selected from a catalogue of $\sim 19000$ galaxies 
with redshift between $z \sim 0.07$ and $z \sim 0.18$ \citep{zabludoff96}.
The second subsample contains 22 E+As from the nearby clusters
DC2048-52, DC1842-63, DC0329-52, and DC0107-46 \citep{caldwell97} 
at $z \sim 0.05$. 
Seven E+A galaxies from rich clusters at $z \sim 0.31$ 
(AC103 and AC114) constitute the 3rd subsample \citep{couch87}.
Some ``control'' galaxies 
were observed also (the 4th subsample, 13 galaxies). These galaxies have 
well-known properties and provide a reference sample to compare the observables of 
the E+A sample. The control sample includes mostly elliptical
and lenticular galaxies (from clusters and the field), as well 
as a few galaxies between $z \sim 0.01$ to $z \sim 0.04$.
Table \ref{samples} summarizes the sample of 63 galaxies which have been observed.


\section{Observations}

All of the observations presented here were carried out at Las Campanas Observatory, Chile,
during photometric conditions. Most of the images were obtained with 
the 40-inch Swope telescope, using a NICMOS3 HgCdTe array (256 $\times$ 256 pixels, 
0.599 arcsec/pix, $2.5 \times 2.5$ arcmin FOV), in 
March, July, August and November 1998. We employed also 
the 100-inch du Pont telescope in September 1998, with a 
NICMOS3 detector, yielding a scale of 0.42 arcsec/pix ($1.8 \times 1.8$ arcmin FOV). 
We use the following filters: $J$, $H$, and $K$ short ($K_s$), centered 
at 1.24$\mu m$, 1.65$\mu m$, and 2.16$\mu m$, 
respectively, and bandwidths of 0.22$\mu m$, 0.30$\mu m$, and 0.33$\mu m$. 
For a detailed discussion of the photometric system see \citet{persson98}. 

Between 5 and 8 standard stars from \citet{persson98} were observed each night. 
The observation procedure for all the objects (standards included) was as follows.
Each object was observed at several positions on the array,
some amount of time in each position (I call this an {\em observing sequence} 
for a given object in a given filter). The amount
of time depended on the magnitude of the object, on the sky
brightness, and on the linearity regime of the array. The NICMOS3 becomes noticeably
non-linear when the total counts (sky + object) exceed 17000 ADU.
We exposed in each position between 60 and 120 seconds in $J$, 
between 30 and 60 seconds in $H$, and
between 30 and 50 seconds in $K_s$. Total exposure times for a given observing 
sequence and filter varied between 10 and 45 minutes. For the standard stars (with magnitudes 
between $K_s \approx 10 - 12$), the exposure time ranged between 
5 and 20 seconds in each position, for both telescopes. 
Typically, the number of non-redundant 
positions at which each object was observed varied between 4 and 10, depending 
on the size and/or magnitude of the object. 
A pre-reduction was made at the beginning of each observing sequence
in order to estimate the total exposure time required to reach a minimum
central S/N $\approx 12-15$. This pre-reduction consisted in the construction of a 
sky image averaging all of the stacked images for a given object (with 
a sigma-clipping rejection threshold), the subtraction of the sky 
image from each individual image, and the combination after registration
of the individual sky-subtracted images. 

\section{Data Reduction}

The images have to be corrected for all of the instrumental effects, namely,
non-linear deviations, dark-current contribution (dark subtraction),
and pixel-to-pixel response differences (flat-field division).

For the determination of the linearity corrections, we 
take several dome-flats with different exposure time. 
Then, we plot the ratio of the average counts and the integration time of each
frame, as a function of the average counts. After normalizing, we 
transform the count value I$_{in}$ of each pixel into 
I$_{out}$ = I$_{in}$ [1 + C$\times$I$_{in}$], and we solve for the {\em constant} C which 
proved to vary between $1.0 \times 10^{-6}$ and 
$5.0 \times 10^{-6}$. At 14000 counts, for example, which corresponds to 
$\sim 75\%$ of the whole range of the signal, the departure from linearity is 
only 2\%.

At the beginning of the night we took series of 20 dark frames, the
number of series depending on the number of different exposure times we
used through the night.
The flat-field images were constructed from (a) dome-flats and twilight 
sky-flats and (b) the raw science images 
of the night, which allowed us to construct {\em super-flats} from the combination of 
the individual frames.
The useful images for these super-flats were those where the 
objects were faint and/or a small number of objects were observed. 
Because galaxy fields were
in general uncrowded, it was always possible to construct super-flats.
Typically, each super-flat was constructed from no less than 30 science 
images, for each filter. 
The results show that the super-flats allow to correct for fringes appearing 
in the stacked and combined images. The fringes are present in the 3 filters ($J$, $H$ 
and $K_s$), but are particularly prominent in the H images. Except for the presence of
fringes, the dome- and sky-flats are quite similar to the super-flats (at the
0.6\% level). However,
given the better photon statistics of the super-flats and the fact that 
the fringes are better represented by the super-flats
we chose the super-flats to remove the pixel-to-pixel variations. 
From the object frames, the dark and flat-fielding corrections were made 
on the images using the SQIID\footnote{Simultaneous Quad-color Infrared Imaging
Device software package, developed by Michael Merrill and John Mac Kenty.} reduction package,
implemented under IRAF\footnote{IRAF is distributed by the National Optical Astronomy
Observatories, which are operated by the Association of Universities 
for Research in Astronomy, Inc., under cooperative agreement with
the National Science Foundation.}. 
At this stage, a mask with the bad pixels was created (using the dark images)
to correct those pixels by interpolation with their neighbor pixels 
in all the images. 

Once the linearity corrections were done, as well as the dark correction
and flat-fielding procedures in all the raw images, the 
procedure to obtain stacked and combined sky-subtracted images began. This
part was done using DIMSUM\footnote{DIMSUM is the Deep Infrared Mosaicing Software
package developed by Peter Eisenhardt, Mark Dickinson, Adam Stanford, and John Ward, and
is available via ftp from {\tt ftp://iraf.noao.edu/iraf/contrib/dimsumV2/dimsum.tar.Z}}, 
also implemented under IRAF. The general procedure for a 
given {\em observing sequence} follows.
\begin{enumerate}
\item[(a)] A scaling  factor for each image was computed using a 5$\sigma$ iterated
rejection method about the mean.  The scaling  factor  is  the  median  of  the
unrejected pixels, and is stored as a descriptor in the image header. 
This provides a first estimate for the sky level. 
\item[(b)] For each image in the observing sequence, a specified number of neighboring images 
of the sequence were selected. To construct a sky image, we selected only the 
neighboring images taken within $\pm 5$ min from the given image.
This provides a sample of sky images within a short-period of time 
during which the variations of the sky level are not larger than 1\% to 3\% 
of the mean, in order to reduce the r.m.s. variation in 
the thermal emission background as well as the OH sky lines, 
characteristic of the NIR (see Figure \ref{sky}). Furthermore, 
the higher background in $K_s$ produces higher shot noise
even if this background did not vary with time. This is a key step.


At each pixel a specified  number
of  low and high values in the scaled images were rejected and the
average of the remainder values was taken as the sky for that pixel. The resulting sky
image was subtracted from the object image to create a sky subtracted object
image. 
\item[(c)] Cosmic  rays were found using a threshold algorithm applied to the
ratio of the image and a median  filtered  image. The detected cosmic rays
were replaced by the local median. A  cosmic  ray  mask  was
created to record  the  location  of the cosmic rays. 
\item[(d)] For a given observing sequence, a shift list is created to define 
the offsets between the images. To create this
list it is first necessary to have at least one object in common among 
all of the images of the observing sequence (usually a star). 
Next we selected a set of additional objects to improve the determination
of the relative shifts. These objects were used
in constructing the shift list using a centroid-based algorithm. 
\item[(e)] A registration stage
was done by shifting and combining the images of the sequence.
A matching exposure map was also created, which allows to obtain a final
mosaic properly weighted by the effective exposure time of each section
of the mosaic. 
\item[(f)] To provide combined images free of ``holes''
arising from the sky subtraction, two different masks are created for 
each registered image of the observing sequence. The detailed procedure is explained 
in the DIMSUM package. 
\item[(g)] The sky subtraction
is repeated as in the first pass, before the masking procedure, 
except that the pixels in the individual masks derived from (f) are ignored. 
\item[(h)] Finally, all the sky-subtracted images of an observing 
sequence are shifted and combined, and the different parts of the mosaic 
are scaled to an exposure time of 1 s. 
\end{enumerate}

\section{Photometry}

\subsection{Photometric calibrations}

The photometric calibrations were done using the faint standard stars
from the list of \citet{persson98}. This list includes 
standard magnitudes in $J$, $H$, $K$ and $K_s$ for equatorial and southern 
photometric standard stars. Five to eight standards were observed 
every night at airmasses similar to those of the galaxies (no larger than 1.2). 
This minimizes the dimming and reddening due to the airmass contribution, especially in 
colors involving the $J$ filter. 

Instrumental magnitudes were computed using the code SExtractor 
\citep{bertin96}, which computes isophotal, isophotal corrected,
and total magnitudes for all the objects detected above a given 
threshold. We have also computed instrumental aperture magnitudes using 
DAOPHOT, and verified that DAOPHOT
aperture magnitudes for the standards do not differ by more than $0.5\%$ from the total
magnitudes yielded by SExtractor. The aperture used for 
the standards in DAOPHOT is the maximum aperture after 
analyzing the shape of the grow curve for the instrumental magnitudes, and 
typically take radii values $\sim 6.0$ arcsec. We conclude that the instrumental magnitudes 
given by SExtractor are reliable, which we adopt hereafter. 

The adopted photometric transformations between the instrumental and the calibrated 
magnitudes are:
\begin{eqnarray}
\label{phot}
J & = & A_1 + j - 0.10 X \\ 
H & = & A_2 + h - 0.04 X \\ 
K_s & = & A_3 + k_s - 0.08 X 
\end{eqnarray}
where the $A_N$ coefficients ($N$ = 1, 2, 3) are the zero points,
$X$ is the airmass, and the extinction coefficients are from \citet{persson98}. 
Note that we do not try to solve for airmass corrections night by night, as this
can lead to spurious values for coefficients if the extinction is variable 
and non-gray. The latter is relevant at the filter passband edges where water
vapor influences the effective width of the passband \citep{persson98}. 

The zero points $A_1$, $A_2$ and $A_3$ were
determined on a nightly basis, and proved to vary between $1\%$ and $7\%$. 
We do not include color terms in these transformations (equations 1, 2, and 3)
since they are smaller than 0.04 mag, a value close to the observational
magnitude errors. We emphasize
that all of the standards and the galaxies reported in this paper
were observed during completely photometric nights. The photometric transformations
have typical r.m.s. residuals of $\sim 0.02 - 0.05$ mag on 
both telescopes (see Figure \ref{stds_error}). This gives an internal error
in the photometric calibrations around 2\% to 5\%.


The main source of error are, in fact, the short-term sky fluctuations (in particular
in $K_s$), which are of the order of 3\% to 6\% in time intervals spanning
the longest exposure time of individual frames during each of the observing sequences 
(120 s in $J$, 60 s in $H$, and 50 s in $K_s$). 
Some galaxies were observed twice, even using the two telescopes, and therefore
there is a good estimate of the global photometric errors, which proves to be around 
7\% for photometric nights. The observation of the same object during two 
photometric nights but with different telescope/instrument is the best 
way to estimate the photometric quality of the data (see \S 5.2), and the 
real dispersion of magnitudes.

\subsection{Galaxy photometry}

Instrumental apparent magnitudes for all the galaxies 
were obtained on the registered and 
combined images using the code SExtractor \citep{bertin96}. 
Given the differences in size, shape
and luminosity of the galaxies, the total magnitude is a better 
estimator than the aperture or isophotal magnitudes. 
SExtractor computes aperture magnitudes, isophotal magnitudes
and ``total'' magnitudes for all of the objects detected above a given threshold. 
The total apparent instrumental magnitude for a given object is given by
one of the two following approaches. (1) It is computed using an {\em adaptive}
aperture magnitude or (2), using a {\em corrected} isophotal magnitude. 
In order to give the best estimate of the total magnitude, the adaptive 
aperture method is performed, except if a neighbor is suspected to 
bias the magnitude by more than 0.1 mag. If this happens, the 
corrected isophotal magnitude is taken as the total magnitude. 
This leads to the so-called MAG\_BEST magnitude, in the 
SExtractor output catalogue.

In order to check the calibrated magnitudes for our galaxies, and
to have an idea of the accuracy of our total magnitudes, we
observed some of the galaxies on two different nights, with the 
40-inch and the 100-inch telescopes.
For the 8 galaxies which were observed twice (2 for each subsample), 
we found r.m.s differences $\Delta J = 0.037$, 
$\Delta H = 0.042$ and $\Delta K_s = 0.061$. These differences, 
although large when taken at face value, represent the most 
realistic errors in the total magnitudes, 
since they were obtained with different instrumental set-ups during
different observing runs.

\subsection{Apparent magnitudes and K-corrections}

Once the instrumental magnitudes are calculated using SExtractor, they
are transformed to the standard system using the package PHOTCAL in IRAF. 
Table \ref{all_calib} shows the apparent
calibrated total magnitudes for all of the galaxies of the 4 subsamples. 
The $K_s$ magnitude for galaxy \# 25, AC114\_89, was not computed since it was
observed under possibly non-photometric conditions, and 
was marked with a NC (not calibrated). 
No internal reddening correction was applied to these magnitudes, nor a Galactic
foreground extinction correction: both corrections are smaller than the photometric 
error and, in particular, are smaller than the uncertainty given by the 
K-correction, as 
we show in \S 5.6. The reddening by dust is $\Delta(J - H) \lesssim 0.03$ and 
$\Delta(H - K_s) \lesssim 0.02$, if we consider a simple screen model 
based on the reddening law of \citet{cardelli89}. If we consider a more
complicated extinction model, following the star-dust mixture recipe by
\citet{wise96}, the amount of reddening is similar. The correction 
due to Galactic extinction is also small for 
the 3 passbands ($\lesssim 0.03$), which proves to be well within the photometric 
uncertainties (for the Galactic reddening corrections in the 
near-IR photometric bands see \citet{schlegel98}). We emphasize, however, that 
Galactic and internal reddening corrections are systematic effects, while the 
photometric uncertainty is random. 
Given their small values, no attempt is made to correct for the extinctions. 
If we include foreground and internal extinction, 
the $J - H$ color redden probably no more than 0.03 -- 0.05 
(but see discussion at the end of \S 6). 


Since the K-terms can significantly modify the intrinsic colors of the 
galaxies, they are critical in correcting the observed colors and magnitudes
to the galaxy rest-frame. If $m_1$ and $m_2$ are the apparent magnitudes 
in the passbands 1 and 2, respectively, for a galaxy at a redshift $z$ and with a 
known spectral type $T$, then the rest-frame color for this galaxy is
\begin{equation}
M_1 - M_2 = m_1 - m_2 - \left\{K_1(z,T) - K_2(z,T)\right\},
\label{rest_col}
\end{equation}
where $M_1$ and $M_2$ are the corresponding absolute magnitudes,
$K_1(z,T)$ and $K_2(z,T)$ are the K-corrections for the passband
1 and 2, respectively, for the galaxy with spectral type $T$ at redshift $z$.
K-corrections are not included in the magnitudes and colors presented in Table
\ref{all_calib} since they can have a wide range of values depending on 
the spectral energy distribution (SED) employed in their computation.
When the SED is not available for a given object, it is common practice (although risky) 
to adopt K-terms from the correlation between spectral type
and morphological classification, provided the latter is available.

In our case, we only have approximate morphological types for the E+As
from the literature, nor do we have spectral information for the galaxies
in the near-infrared part of the spectra. Even though most of our E+A galaxies 
are early types, we can ask the following. How will the near-IR 
K-corrections depend on the spectrophotometric model of galaxy evolution used
to compute them? In order to study the model dependence in $J$, $H$ and
$K$, take for example 2 SED models which 
provide (or allow us to compute) K-terms in the near-IR.
The first K-terms were taken directly from \citet{poggianti97}, who 
computed K-corrections from the 
near-UV to the infrared. \citet{poggianti97} provides K-corrections in several bands in the 
Johnson-Bessel \& Brett photometric system, up to $z = 3$ as a 
function of morphological type. The values are computed according 
to an evolutionary synthesis model that reproduces the integrated 
galaxy spectrum in the range 1000-25000 \AA, and uses the code of 
GISSEL93 \citep{bruzual93}. The models are instantaneous bursts with 
solar metallicity and Scalo IMF \citep{scalo86}. The age after the 
burst gives the SED which is compared with galaxies of known morphological
type through colors. Note that the comparison is done in the optical part of the 
spectrum, mostly between 3000 and 8000 \AA. 
The second set of K-corrections were derived using the model PEGASE\footnote{Projet 
d'Etude des GAlaxies par Synth\`ese Evolutive.}
\citep{fioc97} to generate synthetic spectra between 7000 \AA~
and 30000 \AA, and convolving these SEDs with the filter response functions
(see Persson \etal 1998), using the definition of the K-correction \citep{oke68}. 

Figure \ref{K_poggianti} shows the K-corrections in $J$, $H$ and $K$, calculated by 
\citet{poggianti97} for 3 different morphological types, namely, 
E (solid line), Sa (dotted line) and Sc (dashed line).
Note that the K-corrections in the near-IR are not necessarily
small. Nevertheless, in most of
the photometric bands they do not depend strongly on the spectral type
or the morphological type. 
K-corrections are large (and negative) for the 
$K$ band, for $z \lesssim 0.5$, for all galaxy types (this makes galaxies to appear
brighter than they really are). The average 
redshift in our sample is $\approx$ 0.08, and K-corrections 
in all the bands are less than 0.1 mag for most of the cases. The 
exceptions are the E+A galaxies in AC103 and AC104 (at $z \sim 0.3$). 
For these objects K-corrections can be larger and around $-0.3$ mag in $K$
for the late type galaxies. 
Figure \ref{K_pegase} shows the K-term calculated from PEGASE. 
These K-corrections are calculated from SEDs with solar metallicity
and also instantaneous bursts. The IMF is from \citet{scalo86}.
In PEGASE, the authors define their morphological types by directly comparing 
spectra generated from their models with \citet{kennicutt92} optical 
spectra of nearby galaxies. \citet{poggianti97}, on the other hand, matches 
colors obtained from her model with observed colors of galaxies, taken 
from \citet{persson79} and \citet{bower92a, bower92b}.


Comparing Figure \ref{K_poggianti} and \ref{K_pegase}, we conclude that although
the K-corrections are quite different from one model to another, they are similar
for $z \lesssim 0.1$. For $z \gtrsim 0.2$, differences are larger. 
K-corrections for different Hubble types are more similar if they are derived 
from PEGASE than from the models of Poggianti. Figure \ref{K_diff} shows the
differences between these K-corrections for the two models, in the 3 passbands,
and for the 3 Hubble types. Up to $z \sim 0.5$ the difference for the E type
in $J$ and $H$ is less than 0.05 mag. However, the difference is $\sim 
0.1$ mag for the later types. Equation (\ref{rest_col}) implies that the 
differences in $J - H$ will be less than 0.05 mag. However, this is not
the case for colors involving the $K$ band ($J - K$ and $H - K$), due 
to the large difference in the K-corrections, for all the Hubble types, 
as shown also in Figure \ref{K_diff}. The difference in $K$ for the K-corrections
reaches values $\sim 0.4$ mag at $z \sim 0.3$. This shows that K-correction 
uncertainty will have the largest impact on rest-frame colors. 
Other studies also show large differences between their K-corrections, although some 
of them are comparable to the values of this work, showing also large negative K-corrections
in $K$ \citep{frogel78, persson79, bershady95}. For example, \citet{bershady95} 
obtains type-averaged K-corrections in $K$,
reaching $-0.33$ and $-0.60$ at $z = 0.14$ and $z = 0.30$, respectively. These values
are larger than values from \citet{poggianti97}, but similar to those obtained 
from PEGASE (see Figure \ref{K_poggianti} and \ref{K_pegase}). 


\section{Comparison with models}

As shown in the preceding section, K-corrections in the near-IR can be 
very different depending on the spectrophotometric models used. Therefore,
we do not use rest-frame colors, i.e. we do not de-redshift the data. 
Instead, we {\em redshift} current epoch SEDs. Although this approach is similar
to work with rest-frame colors, it is more robust, since the SEDs of the current 
epoch models can be determined absolutely. In order to give an idea whether
synthetic SEDs compare well with spectra of galaxies at the current 
epoch, we consider GISSEL96 models \citep{charlot96}
and compare them with real, local galaxy spectra of known morphological types, 
given by \citet{kennicutt92}. As it is well-known, the 
age-metallicity degeneracy prevents us for deriving age and metallicity directly 
from colors, as was shown by \citet{worthey94, ferreras98}. Therefore, we consider 
instantaneous bursts of fixed (solar) metallicity. Subsequent evolution 
is determined by adopting passive stellar evolution, measured in Gyrs and 
indicated by the label ``age'' for each model spectrum in Figure \ref{spectra}. 
A simple $\chi^2$ test is used to determine the model spectra closest to the 
observed (Kennicutt) sample. We use a starting sample of 20 GISSEL96 spectra
and 27 spectra representative of normal galaxies of known Hubble types 
\citep{galaz98}. Figure \ref{spectra} shows the better spectral match between some 
Kennicutt spectra and the 20 selected GISSEL96 models.


The Hubble sequence fits well with an evolutionary sequence in the optical,
but care has to be taken in the interpretation since 
more than one solution can be obtained from a synthetic set where both age and
metallicity vary \citep{ronen99}. Even though metallicity can vary from one galaxy
to another, it is realistic to set metallicity close to solar. Extremely metal-poor
(Z $\lesssim 0.5$ Z$_\odot$) or metal-rich (Z $\gtrsim 1.5$ Z$_\odot$) cases are unlikely
in this set of galaxies \citep{liu96}. Moreover, the fact that colors are obtained 
from {\em integrated} total apparent magnitudes, imply that colors are an average
over the whole galaxy light and therefore likely to be representative of
solar metallicity or lower in the luminosity weighted mean (see for
example \citet{edmunds92}). 

In order to compare the observed colors with models, we take the 20 spectra from 
GISSEL96 and we ``redshift'' them to several redshift values (from the rest-frame to 
$z = 0.5$). Afterwards, we compute synthetic colors using $J$, $H$ and $K_s$ passbands
\citep{persson98} for the 20 synthetic spectra. 
Figure \ref{obs_colors} shows the color-color diagram for the E+A sample and the 
control sample (indicated as filled circles), as well as for the model spectra 
(indicated as open symbols) situated at different redshifts (as indicated by labels). 
We include 3 different evolving tracks in figure \ref{obs_colors}, for instantaneous
bursts after 1 Gyr, 3 Gyr and 16 Gyr indicated by circles, squares, and triangles,
respectively. After 10 Gyr, the near-IR colors are almost 
independent of age, for a given redshift. 

Figure \ref{obs_colors} shows that there is an overall agreement between near-IR 
colors of all subsamples and models, except for subsample 3. The average 
color $<H - K_s> = 0.41$ $(\sigma = 0.05)$ of subsample 1 (average redshift 
$<z> = 0.09$, $\sigma = 0.02$) agrees well with any model older than 3 Gyr at $z = 0.10$.
However, the average $<J - H> = 0.66$ $(\sigma = 0.06)$ appears bluer than the same
models by $\sim 0.1$ mag. Otherwise, $<J - H>$  is well fitted by a 
model with age $\lesssim 3$ Gyr at $z = 0.10$, but then $<H - K_s>$ of 
subsample 1 is redder by $\sim 0.1$ mag. These differences are twice the
color dispersion for this subsample. Therefore we can conclude that colors
of the models and the data do not differ by more than $2\sigma$. 
In subsample 2, the average
colors $<H - K_s> = 0.29$ $(\sigma = 0.07)$ and 
$<J - H> = 0.69$ $(\sigma = 0.05)$, with average redshift $<z> = 0.046$ $(\sigma = 0.014)$, 
are well fitted by a model at $z = 0.05$ and 2.8 Gyr. Subsample 3, having
$<H - K_s> = 0.61$ $(\sigma = 0.23)$, 
$<J - H> = 0.75$ $(\sigma = 0.25)$, and average redshift $<z> = 0.31$ $(\sigma = 0.01)$
is not fitted by the GISSEL96 models, even though the average color $<H - K_s>$
is closer to the $z = 0.3$ redshifted color of models. Subsample 4 (the 
control sample) matches the models colors well, despite the rather large scatter. 
This subsample has average colors  $<H - K_s> = 0.29$ $\sigma = 0.08)$,  
$<J - H> = 0.74 (\sigma = 0.06)$, and average redshift $<z> = 0.030$ $(\sigma = 0.012)$.
These average colors correspond to a model located at $z = 0.05$ and age 3 Gyr.
This subsample shows a larger scatter in the color-color diagram. 
Most of these galaxies are nearby
galaxies (from the PGC and NGC catalogues) and some 
galaxies from DC clusters \citep{caldwell97}. All have
secure Hubble types, and most of them have known photometric properties in the
optical (for $B$ and $R$ total magnitudes see Table \ref{all_calib}). 
The majority of these galaxies are well matched by the 
colors provided by the spectrophotometric models, for ages representative of
early type galaxies. These
galaxies have large apparent radii, and therefore, their
photometry is more sensitive to color gradients. This is not a problem for more
distant galaxies because of the poorer spatial resolution. 


Now we compare color properties of subsample 1 (field E+As from the LCRS) 
and 2 (cluster E+As). From Figure \ref{obs_colors} it is apparent that subsample 1
has the same average $<J - H>$ (with a difference of 0.03), but a 
redder $<H - K_s>$ than subsample 2 (see above). The difference of
0.12 mag is $2.4\sigma$ and $\sim 1.7\sigma$ away from the 
intrinsic dispersion of subsample 1 and subsample 2, respectively. 
The expected color difference due to K-corrections between $<z> = 0.09$ (subsample 1)
and $<z> = 0.046$ (subsample 2) is $\sim 0.06$ mag, 
for a 2 Gyr model 
(half of the 0.12 color difference between the two subsamples), which fits
the average colors of both subsamples 1 and 2 better. Therefore, we can only 
conclude with a $\sim 1.5\sigma$ confidence level that E+A galaxies from 
the field are redder than cluster E+As. The fact that dust extinction is much more 
notorious in $J - H$ than in $H - K_s$ suggest that the color difference
observed in $H - K_s$ between subsample 1 and subsample 2 is not due to 
differential internal dust extinction. However, because of the observed 
color dispersion, we cannot give a robust answer
supporting stellar population differences instead of internal reddening 
differences due to extinction. We stress
that our differences are only at $1.5\sigma$ significance level. 
It is worth noting that $J - H$ color would {\em redden systematically} 
$\sim 0.03 - 0.05$  if we account for foreground or internal extinction
(see \S 5.3).
This would make ages inferred from colors (see Figure \ref{obs_colors})
slightly older (0.5 to 1 Gyr), but in any case alter the results of 
the analysis, since changes are the same for all the galaxy samples. 

\section{Further analysis and discussion}

\subsection{Photometry uncertainties}

In order to interpret correctly the color properties
of the observed galaxy sample, it is important to keep in mind
the sources of uncertainty which affect the colors. 
The first source of uncertainty is of course the 
data acquisition itself. Given the nature of the near-IR imaging,
the thermal variation of the sky affects the photometry for the faint
objects, which require longer integration time than the brighter ones, 
sometimes much longer than the typical time of the 
sky fluctuations (see Figure \ref{sky}). However, the nature of these variations is well
understood and the fact that the sky fluctuations are sampled in {\em real-time} 
and subtracted for each image can reduce this error to 5\% 
(see \S 3). The second important source of errors is the procedure employed to compute the
magnitude. It is well known that total magnitudes depend on the cut
level where the light contribution is null or
not significant. In our case, SExtractor computes total magnitudes
integrating all the light up to a given threshold above the sky 
(typically 1.5$\sigma$), and fitting elliptical isophotes to the profiles.
An  elliptical aperture for a given galaxy, defined by 
the elongation $\epsilon$ and position angle $\theta$, is computed 
from the 2nd order moment in the light distribution, above the 
isophotal threshold. The ``first moment'' $r_1$ is then 
computed\footnote{$r_1$ is defined as $r_1 = \frac{\sum_r r I(r)}{\sum_r I(r)}$.} 
within an aperture twice as large as the isophotal aperture, in 
order to reach the light distribution in the wings. This approach is very
similar to the approach of \citet{kron80}. The parameter $r_1$ is then used to define 
the adaptive aperture where the total magnitude will be computed. The 
main axes of the ellipse are defined as $\epsilon k r_1$ and
$k r_1 / \epsilon$, where $k$ is a value to be fixed by the user. We
carried out some tests with both faint and bright galaxies and found that the
value $k = 2.5$ allows us to include between $90\%$ and $95\%$
of the total flux without introducing additional noise within the 
aperture. Further details can be found in \citet{arnouts96}. This
procedure ensures that not more than 5\% of the light is lost.

Another source of uncertainty is the photometric errors due to the 
transformation of the instrumental magnitudes to calibrated 
magnitudes. This process is well understood and in general the 
scatter is small. The errors of the zero points are $\sim 2\%$ to
$\sim 7\%$. 

The largest uncertainties (now for rest-frame colors) 
are due to the K-corrections. These
uncertainties, as shown in the previous section, can be very large
for galaxies with $z \gtrsim 0.25$, where the change in magnitude
produced by the computation of K-corrections assuming one or another
SED can reach differences as large as 30\%, 
propagating these differences to the rest-frame colors (see Figure \ref{K_diff}). 
For galaxies with $z \lesssim 0.2$, differences
are smaller: $\sim 10\%$ for $0.15 \lesssim z \lesssim 0.2$ and 
$\sim 5\%$ for $0 \le z \lesssim 0.15$. In order to compute reliable K-corrections,
it is fundamental to obtain calibrated spectra at 9000 \AA $\lesssim \lambda
\lesssim$ 25000 \AA~ for different spectral types, including E+A galaxies.
Of course, the nature of the 
uncertainties lies in the fact that K-corrections are expressed in 
term of the morphological type instead of the spectral type. The
morphological type relies on a subjective classification procedure, often
dependent on the passband through which the images are obtained (more or less 
sensitive to the star population which delineates the galaxy
morphology) and is always strongly dependent on the image quality. On the other
hand, there is no unique and reliable relationship between the spectral
type and the morphological type of the galaxies. Even though this is 
approximately true for normal Hubble types \citep{folkes96, galaz98},
the dispersion can be large for some spectral types 
or active galaxies \citep{sodre99}, leading to large uncertainties in the
K-correction = $f(z$, T-type). However, we note that independently of what
spectrophotometric models are used in obtaining rest-frame colors,
the K-corrections in $K$ (or $K_s$ band), are only {\em weakly}
dependent on the spectral type for $z \lesssim 0.2$ (see Figures 
\ref{K_poggianti}, \ref{K_pegase}, and \ref{K_diff}).

\subsection{Implications from near-IR colors}

We now examine some color properties of the E+A galaxies observed 
in the near-IR, keeping in mind the limitations of the accuracy
of our photometry, as discussed above.

Studying the position of the E+A galaxies in the $H - K_s$/$J - H$
plane shown in Figure \ref{obs_colors}, we see that {\em field}
galaxies located at $<z> \sim 0.09$ (subsample 1), have an average
$J - H$ color similar to that of E+A galaxies located in nearby clusters 
($<z> \sim 0.05$, subsample 2), but are slightly {\em redder} in the 
average $H - K_s$ color (see preceding section). The fact that the color
difference of 0.12 mag in only significant at $\sim 1.5\sigma$ level 
prevents us from proposing a robust conclusion. However, we can now ask 
how the K-corrections can change this result. Here we examine the answer
to this question using the two sets of K-corrections show in \S 5.3
the PEGASE and the Poggianti K-corrections. 

Figure \ref{color_average} shows average rest-frame colors for our sample
of galaxies computed using both sets of K-corrections. Also shown are the colors
of the sample of elliptical galaxies from \cite{silva98}. We show the average
colors for the {\em cluster} and the {\em field} galaxies separately. This 
Figure demonstrates that, although the color differences are small, the same 
trend is observed, independently of which set of K-corrections is used. The color
difference in $<H - K_s>$ between field and cluster E+As is 
about 0.04 mag using PEGASE K-corrections and 0.15 mag
using Poggianti K-corrections. Note that in 
Figure \ref{color_average} we compare cluster-field colors also for the 
LCRS sample (3 LCRS E+As belong to clusters). The field E+As from LCRS are also
redder in $<H - K_s>$ than the LCRS cluster E+As. 


As demonstrated by \citet{persson83}, stellar
populations containing a large fraction of AGB stars (1 to 3
Gyr old), have redder $H - K$ color (but similar $J - H$ index), compared
with populations that lack such stars. This might suggest that the 
E+A galaxies in the field have a larger fractions of AGB stars than those in 
clusters. We emphasize that, although the difference given by the K-correction between 
$z \sim 0.1$ and $z \sim 0.05$ for subsamples 1 and 2, respectively, does
change the corresponding average colors, the observed color trend field/cluster
does not change. 

Note that three out 21 LCRS E+As are embedded in clusters (LCRS \# 4, 11 and 20, 
These galaxies have an average color $<H - K_s> = 0.160 \pm 0.041$,
using the PEGASE K-corrections, and $<H - K_s> = 0.260 \pm 0.005$, using the Poggianti
K-corrections. These values are $35\%$ and $22\%$ bluer, respectively, than 
the average $H - K_s$ color for the LCRS E+A galaxies located in the field, and 
are consistent with the comparison field/cluster between subsamples 1 and 2. 


Galaxies in more distant clusters (subsample 3), 
appear redder in $J - H$ (at 2$\sigma$ significance level) 
than those at lower redshift (compared with both
subsamples 1 and 2). As discussed above, although for this subsample 
K-corrections are critical, the $J - H$ color does not change if one uses
a different set of K-corrections. This could be interpreted as a temperature
change of the first-ascent giant branch (FAGB) in the 
stellar populations of these $z \sim 0.3$ E+A galaxies (see \citet{charlot96}). A
further spectroscopic analysis in the near-IR would settle this question, 
and also will help to disentangle possible significant extinction in the 
$J$ band.

One can also compare the integrated rest-frame 
colors between the E+A galaxies from subsample 2 with the control galaxies which 
{\em also} belong to these nearby clusters (e.g. galaxies \# 52, 53, 59, 
60 and 61 in Table \ref{all_calib}.
We note that the average $H - K_s$ color
for both sets of galaxies is similar, and therefore any difference (in the mean)
is observed between E+A galaxies and elliptical galaxies belonging to the 
same cluster.
This is not the case if one compares the 
colors between subsamples 1 and 2, as shown before. 
We emphasize that the average $J - H$ color of the E+A galaxies of subsamples 1 and 2
is similar to the average $J - H$ of the control sample 
(at $1\sigma$ significance level, see Figure \ref{obs_colors}). 

It is worth noting that all the K-corrections used to obtain average rest-frame colors, 
as shown in Figure \ref{color_average}, have been computed using solar
metallicity models, assuming that K-corrections, for a given IMF, age
and SFR scenario do not depend strongly on metallicity. We tested this
assumption using GISSEL96 SEDs with different
metallicity. Several tests were carried out for different ages, 
IMFs, and SFRs, and metallicity between the extreme values of 
[Fe/H] $=-1.65$ and [Fe/H] $= +1.00$. 
Differences in K-corrections between these two extreme metal-poor and metal-rich models 
can reach up to 0.3 mag in $J$ at $z = 0.3$, for a large range of fundamental 
parameters (age, IMF, and SFR). For more modest metallicity differences 
between models (probably more realistic), variations in K-corrections, for the different near-IR
photometric bands, are between 0.15 mag for $J$ and 0.05 mag for $H$ and $K_s$, at $z = 0.3$. 
For smaller redshifts, these differences are even smaller.  
Figure \ref{diff_metal} shows K-correction differences in near-IR bands, as a function of redshift,
for two SEDs (shown in the inset) with different metallicity ([Fe/H] $= -0.30$ and 
[Fe/H] $= +0.1$), derived from instantaneous bursts with 
the same age and IMF (in both cases Scalo IMF). These K-corrections differences 
imply $J - H$ and $H - K_s$ colors shifts no larger than 0.08 mag up to $z \sim 0.3$, given 
that differences in K-corrections, due to different 
metallicity have the same sign. We conclude that for a typical interval of 
metallicity found in the field and in clusters, the effect of varying metallicity should 
not be significant on the K-correction uncertainties, and hence on rest-frame colors. 
However, for more accurate estimates of near-IR colors from broad band photometry, 
especially at higher redshift ($z \gtrsim 0.5$), metallicity does play a 
significant role on the K-corrections. 


\section{Summary and conclusions}

The E+A galaxies reported here include 32 galaxies
from clusters and 18 galaxies from the field. In addition, 13 nearby galaxies
which do not present post-starburst activity, were observed (5 located in clusters
at $z \sim 0.05$ and 8 located in the field at very low redshift). All the galaxies
have been observed in the near-IR bands $J$, $H$ and $K_s$ during 
photometric nights at Las Campanas Observatory.
Total apparent magnitudes and colors were derived. The color-color 
diagram $H - K_s$/$J - H$  of the observed galaxies is compared to
the expected corresponding colors of spectrophotometric models of galaxy 
evolution, at different redshifts. The models are those generated by
GISSEL96 \citep{charlot96}. There is an overall agreement between these
expected colors and the observed ones, for the E+A located in nearby 
clusters ($<z> \sim 0.05$) and for E+As located in the field ($<z> \sim 0.1$).
The comparison of the colors of these two samples shows that even though 
cluster E+As appear bluer than field E+As, the color difference is only significant at
$\sim 1.5\sigma$ level, and therefore we cannot strongly affirm that stellar 
population differences are observed between these two populations. 

The colors of the E+A galaxies located in more distant clusters $z = 0.3$, 
on the other hand, do not agree with the color expected from models. In the 
mean, they appear bluer than expected in $H - K_s$ (by $\sim 0.3$mag) 
and redder in $J - H$ (by $\sim 0.15$ mag). The possible interpretation of
the failure is strong internal reddening (mostly in the $J$ band), not 
considered in models. 

In order to derive a more complete comparison with models, 
rest-frame colors were also obtained using two different sources of K-corrections:
one based on the work of \citet{poggianti97}, and the other computed using
the spectrophotometric model PEGASE \citep{fioc97}. 
We have shown that such K-corrections can be significant for $z \sim 0.2$ in the
$K$ bands (or any band centered at 2$\mu m$), although they are not a strong function
of spectral type. In addition, large differences
exist in the K-corrections between these two models, having a large impact on
the derived quantities, like rest-frame colors for high redshift galaxies. 
We have compared average rest-frame colors of E+A galaxies located in the 
field and in clusters. Results show that average rest-frame near-IR colors of E+A
galaxies located in clusters at $z \sim 0.05$ \citep{caldwell97} 
and field E+As located at $z \sim 0.1$ (from the LCRS, Zabludoff \etal 1996), 
follow the same color trend in $J - H$ and $H - K_s$ observed in the comoving
color-color diagram: E+A galaxies located in nearby clusters appear 
bluer than field E+As ($z \sim 0.1$). 


As well as comparing the observed colors with the GISSEL96 colors at
different redshifts, the models do not fit the rest-frame colors 
of the E+A galaxies observed in clusters
at $z \sim 0.3$. Their $H -K_s$ colors appear bluer (using the K-corrections 
of PEGASE), or redder (using the K-corrections of Poggianti) compared to
the models. Their $J - H$ color index, although not particularly sensitive to
one or the other K-correction, is also redder than the colors predicted by the 
models. The color of control galaxies, most of them ellipticals at $z \lesssim 0.01$, and
the others from clusters at $z \sim 0.05$, agree with the near-IR colors predicted
by models. 

Integrated colors between the {\em field} E+As and the {\em cluster} 
E+As of the LCRS (LCRS \# 4, 11 and 20; see Table \ref{all_calib}) are
similar, although those in clusters seem to be slightly ($\sim 25\%$) bluer in $H - K_s$ 
than the average color. This result is the same for both sets of rest-frame 
colors, the set corrected by the PEGASE K-corrections and the set corrected 
by the Poggianti K-corrections (see Figure \ref{color_average}). 
On the other hand, the corresponding $J - H$ color is similar for 
the cluster and field E+As in subsample 1. 

In order to build more robust results, more field and cluster 
E+A galaxies have to be observed  between $z = 0.1 - 1.0$. Spectroscopic 
observations of normal and E+A galaxies at different redshifts 
in the near-IR are necessary to (1) obtain calibrated SEDs and 
realistic K-corrections, and (2) to compare the spectra of the E+A galaxies 
with those of normal galaxies in the whole spectral range 
3500 \AA~ $\lesssim \lambda \lesssim 25000$ \AA. We expect to continue this 
research by imaging new E+A galaxies in the near-IR at higher redshift, as
well as obtaining near-IR spectra in order to construct a useful and larger database of 
normal and post-starburst galaxies in a large spectral range. In order to
increase the number of E+A galaxies, some field galaxies already classified
as E+As, are being observed in the near-IR at Las Campanas.
Some of these galaxies 
belong to the ESO-Sculptor Survey \citep{delapparent97}, and results
will be published soon. Other wide-field surveys will provide a wealth of data 
for E+A galaxies at $0.01 \lesssim z \lesssim 0.2$, like SLOAN \citep{loveday98,fan98}, 
and the 2DF survey \citep{colless98}, whose 
data are expected to become available to the public. In a forthcoming paper, 
we shall investigate systematic properties on the surface photometry and 
colors of the E+A galaxies. 

\acknowledgements

I would like to thank the anonymous referee for useful comments and suggestions on how to improve 
this paper. I thank Ron Marzke, Eric Persson, and Ann Zabludoff for fruitful discussions
on the nature of the E+A galaxies. I acknowledge Mauro Giavalisco for his help
with the DIMSUM software. I also acknowledge Mario Hamuy, Ren\'e M\'endez,
Mark Phillips, Miguel Roth, and Bill Kunkel 
in helping to improve the preliminary version of this paper. 
It is a pleasure to thank all the staff at Las Campanas Observatory. 
This research has made use of the NASA/IPAC Extragalactic Database (NED), 
which is operated by the Jet Propulsion Laboratory, California Institute
of Technology, under contract with the National Aeronautics and Space
Administration. This work is made possible through the fellowship \# C-12927, under 
agreement between Fundaci\'on Andes and Carnegie Institution of
Washington.

\clearpage 


\begin{deluxetable}{lllccclcc}
\tablewidth{0pt}
\renewcommand{\baselinestretch}{1.4}
\tablecaption{The sample. \label{samples}}
\tablenum{1}
\tablehead{
\colhead{ID\tablenotemark{(1)}} &
\colhead{S-ID\tablenotemark{(2)}} &
\colhead{Name\tablenotemark{(3)}} &
\colhead{RA\tablenotemark{(4)}} & 
\colhead{DEC\tablenotemark{(5)}} & 
\colhead{$z$\tablenotemark{(6)}} & 
\colhead{Cluster/Field\tablenotemark{(7)}} & 
\colhead{T-type\tablenotemark{(8)}} & 
\colhead{Reference\tablenotemark{(9)}}}
\startdata
\cutinhead{E+A galaxies} 
1 & 1 & g515            & 15:24:26 & +08:09:06 & 0.0870 & Abell 665 & 0 & (1) \\
2 & 1 & dc204852\_26    & 20:49:52 & $-$53:02:58 & 0.0397 & ACO 3716   & $-$2 &  (2) \\
3 & 1 & dc184263\_39m   & 18:42:49 & $-$63:12:28 & 0.0144 & DC1842-63  & $-$3 &  (2) \\
4 & 1 & dc204852\_100   & 20:51:49 & $-$52:44:45 & 0.0493 & ACO 3716   & $-$2 &  (2) \\
5 & 1 & dc204852\_148   & 20:49:13 & $-$52:33:51 & 0.0429 & ACO 3716 & $-$2 &  (2) \\
6 & 1 & dc204852\_39    & 20:50:01 & $-$52:59:56 & 0.0489 & ACO 3716 & $-$2 &  (2) \\
7 & 1 & dc204852\_45    & 20:52:10 & $-$52:56:09 & 0.0484 & ACO 3716 & $-$2 &  (2) \\
8 & 1 & dc204852\_104   & 20:51:07 & $-$52:43:34 & 0.0493 & ACO 3716 & 0  &  (2) \\
9 & 1 & dc204852\_149   & 20:48:30 & $-$52:33:07 & 0.0569 & ACO 3716 & 0  &  (2) \\
10 & 1 & dc204852\_192  & 20:51:56 & $-$52:03:45 & 0.0473 & ACO 3716 & $-$5 &  (2) \\
11 & 1 & dc204852\_77   & 20:52:54 & $-$52:47:28 & 0.0452 & ACO 3716 & $-$2 &  (2) \\
12 & 1 & dc204852\_174  & 20:51:46 & $-$52:16:09 & 0.0448 & ACO 3716 & $-$5 &  (2) \\
13 & 1 & dc204852\_184  & 20:54:00 & $-$52:08:15 & 0.0469 & ACO 3716 & $-$2 &  (2) \\
14 & 1 & dc204852\_216  & 20:49:24 & $-$51:56:56 & 0.0490 & ACO 3716 & $-$2 &  (2) \\
15 & 1 & dc204852\_231  & 20:51:40 & $-$51:45:22 & 0.0459 & ACO 3716 & $-$2 &  (2) \\
16 & 1 & dc032952\_135a & 03:29:31 & $-$52:27:18 & 0.0519 & ACO 3128 & $-$2 &  (2) \\
17 & 1 & dc032952\_156a & 03:31:15 & $-$52:22:28 & 0.0604 & ACO 3128 & $-$2 &  (2) \\
18 & 1 & dc010746\_30b  & 01:10:51 & $-$45:51:52 & 0.0267 & ACO 2877 & $-$5 &  (2) \\
19 & 1 & dc032952\_82a  & 03:31:09 & $-$52:36:49 & 0.0576 & ACO 3128 & $-$5 &  (2) \\
20 & 1 & dc032952\_158b & 03:29:35 & $-$52:39:58 & 0.0500 & ACO 3128 & 0  &  (2) \\
21 & 1 & dc010746\_22m  & 01:08:23 & $-$46:09:09 & 0.0200 & ACO 2877 & 0  &  (2) \\
22 & 1 & dc010746\_45m  & 01:09:07 & $-$45:44:29 & 0.0300 & ACO 2877 & 0  &  (2) \\
23 & 2 & ac103\_132     & 20:57:18 & $-$64:38:48 & 0.3047 & AC 103   & 0  &  (3) \\
24 & 2 & ac114\_22      & 22:58:50 & $-$34:48:13 & 0.3354 & AC 114   & 0  &  (3) \\
25 & 2 & ac114\_89      & 22:58:49 & $-$34:46:57 & 0.3169 & AC 114   & 0  &  (3) \\
26 & 2 & ac103\_03      & 20:56:55 & $-$64:40:11 & 0.3118 & AC 103   & 0  &  (3) \\
27 & 2 & ac103\_106     & 20:56:47 & $-$64:40:56 & 0.3091 & AC 103   & 0  &  (3) \\
28 & 2 & ac103\_280     & 20:57:26 & $-$64:42:11 & 0.3111 & AC 103   & 0  &  (3) \\
29 & 2 & ac103\_145     & 20:57:07 & $-$64:38:29 & 0.3105 & AC 103   & $-$2 &  (3) \\
30 & 3 & lcrs01        & 11:01:19 & $-$12:10:18 & 0.0746 & Field    & 1  &  (4) \\
31 & 3 & lcrs17        & 10:13:52 & $-$02:55:47 & 0.0609 & Field   & 0 & (4) \\
32 & 3 & lcrs21        & 11:15:24 & $-$06:45:13 & 0.0994 & Field   & 0 & (4) \\
33 & 3 & lcrs13        & 11:19:52 & $-$12:52:39 & 0.0957 & Field   & 1 & (4) \\
34 & 3 & lcrs14        & 13:57:01 & $-$12:26:47 & 0.0704 & Field   & 0 & (4) \\
35 & 3 & lcrs12        & 12:05:59 & $-$02:54:32 & 0.0971 & Field   & 1 & (4) \\
36 & 3 & lcrs03        & 12:09:05 & $-$12:22:37 & 0.0810 & Field   & 1 & (4) \\
37 & 3 & lcrs16        & 12:19:55 & $-$06:14:01 & 0.0764 & Field   & 1 & (4) \\
38 & 3 & lcrs15        & 14:40:44 & $-$06:39:54 & 0.1137 & Field   & 0 & (4) \\
39 & 3 & lcrs06        & 11:53:55 & $-$03:10:36 & 0.0884 & Field   & 0 & (4) \\
40 & 3 & lcrs08        & 14:32:03 & $-$12:57:31 & 0.1121 & Field   & $-$2 & (4) \\
41 & 3 & lcrs07        & 22:41:09 & $-$38:34:35 & 0.1141 & Field   & 0 & (4) \\
42 & 3 & lcrs20        & 00:38:44 & $-$38:57:12 & 0.0632 & Cluster & $-$2 & (4) \\
43 & 3 & lcrs18        & 00:22:46 & $-$41:33:37 & 0.0598 & Field   & 0 & (4) \\  
44 & 3 & lcrs05        & 01:58:01 & $-$44:37:14 & 0.1172 & Field   & $-$2 & (4) \\
45 & 3 & lcrs19        & 02:07:49 & $-$45:20:50 & 0.0640 & Field   & 0 & (4) \\
46 & 3 & lcrs11        & 01:14:49 & $-$41:22:30 & 0.1216 & Cluster & 0 & (4) \\
47 & 3 & lcrs02        & 02:17:39 & $-$44:32:47 & 0.0987 & Field   & 2 & (4) \\
48 & 3 & lcrs09        & 01:17:38 & $-$41:24:23 & 0.0651 & Field   & 0 & (4) \\
49 & 3 & lcrs10        & 02:11:43 & $-$44:07:39 & 0.1049 & Field   & 0 & (4) \\
50 & 3 & lcrs04        & 04:00:00 & $-$44:35:16 & 0.1012 & Cluster & 1 & (4) \\ 
\cutinhead{Control galaxies}   
51 & 4 & pgc35435      & 11:30:05 & $-$11:32:47 & 0.0178 & Field   & $-$3 & (5) \\
52 & 4 & dc204852\_116 & 20:51:19 & $-$52:40:41 & 0.0441 & ACO 3716 & $-$5 & (2) \\
53 & 4 & dc204852\_66  & 20:51:45 & $-$52:51:19 & 0.0410 & ACO 3716 & $-$5 & (2) \\
54 & 4 & pgc60102      & 17:20:28 & $-$00:58:46 & 0.0304 & Field & $-$2 & (6) \\
55 & 4 & eso290-IG\_050& 23:06:46 & $-$44:15:06 & 0.0290 & Field & $-$2 & (7) \\
56 & 4 & pgc62615      & 18:57:41 & $-$52:31:46 & 0.0280 & Field & 2 & (8) \\
57 & 4 & pgc57612      & 16:15:04 & $-$60:54:26 & 0.0183 & Field & $-$5 & (9) \\
58 & 4 & ngc6653       & 18:44:39 & $-$73:15:47 & 0.0172 & Field & $-$5 & (9) \\
59 & 4 & dc204852\_115 & 20:51:21 & $-$52:39:17 & 0.0440 & ACO 3716 & $-$5 & (2) \\
60 & 4 & dc204852\_126 & 20:51:44 & $-$52:37:57 & 0.0489 & ACO 3716 & $-$2 & (2) \\
61 & 4 & dc204852\_38  & 20:50:05 & $-$53:00:28 & 0.0454 & ACO 3716 & $-$2 & (2) \\ 
62 & 4 & ngc6328       & 17:23:41 & $-$65:00:37 & 0.0142 & Field & 2 & (6) \\
63 & 4 & pgc62765      & 19:05:59 & $-$42:21:59 & 0.0193 & Field & $-$2 & (6) \\ 
\tablenotetext{(1)}{Correlative number of the galaxy.}
\tablenotetext{(2)}{Sample ID. Sample 1, Nearby cluster E+As; sample 2, distant cluster E+As;
			sample 3, LCRS E+As; sample 4, control galaxies.}
\tablenotetext{(3)}{Galaxy Identification used in this paper.}
\tablenotetext{(4)}{Right ascension in hh:mm:ss (J2000).}
\tablenotetext{(5)}{Declination in \arcdeg:\arcmin:\arcsec (J2000).}
\tablenotetext{(6)}{Redshift.}
\tablenotetext{(7)}{Column indicating whether the galaxy belongs to a 
cluster or to the field.}
\tablenotetext{(8)}{Morphological type in T-type units, from the 
de Vaucouleurs classification system \citep{devaucouleurs76}.}
\tablenotetext{(9)}{Reference where quantities other than magnitudes
have been extracted.}
\tablerefs{
(1) \citet{franx93}; 
(2) \citet{caldwell97};
(3) \citet{couch87};
(4) \citet{zabludoff96};
(5) \citet{fairall92};
(6) \citet{devaucouleurs91};
(7) \citet{loveday96};
(8) \citet{spellman89};
(9) \citet{prugniel98}}
\enddata
\end{deluxetable}

\clearpage

\begin{deluxetable}{llccccccccccc}
\tablewidth{0pt}
\scriptsize
\rotate
\scriptsize
\renewcommand{\baselinestretch}{0.90}
\tablecaption{Apparent magnitudes and colors. \label{all_calib}}
\tablenum{2}
\tablecolumns{13}
\tablehead{
\colhead{ID\tablenotemark{(1)}} &
\colhead{Name\tablenotemark{(2)}} &
\colhead{$J$\tablenotemark{(3)}} & 
\colhead{$H$\tablenotemark{(4)}} & 
\colhead{$K_s$\tablenotemark{(5)}} &
\colhead{$(J - H)$\tablenotemark{(6)}} &
\colhead{$(H - K_s)$\tablenotemark{(7)}} &
\colhead{$(J - K_s)$\tablenotemark{(8)}} &
\colhead{$z$\tablenotemark{(9)}} &
\colhead{$B$\tablenotemark{(10)}} &
\colhead{$R$\tablenotemark{(11)}} &
\colhead{S-ID\tablenotemark{(12)}} &
\colhead{T-Type\tablenotemark{(13)}}}
\startdata
&&&&&&&&&&&& \\
1 & g515          & 13.86 0.02 & 13.11 0.05 & 12.75 0.05 & 0.75 0.05   & 0.36 0.07 & 1.11 0.05  & 0.0870 &   \nodata &  \nodata  & 1  & 0 \\
2 & dc204852\_26   & 13.82 0.03 & 13.16 0.04 & 12.99 0.03 & 0.66 0.05  & 0.17 0.05 & 0.83 0.04  & 0.0397 &   16.99 &  15.38  & 1 & $-$2 \\
3 & dc184263\_39m  & 11.01 0.02 & 10.27 0.06 & 10.05 0.03 & 0.74 0.06  & 0.22 0.07 & 0.96 0.04  & 0.0144 &   \nodata &  \nodata  & 1 & $-$3 \\
4 & dc204852\_100  & 14.62 0.04 & 13.95 0.06 & 13.64 0.03 & 0.67 0.07  & 0.31 0.07 & 0.98 0.05  & 0.0493 &   17.61 &  16.19  & 1 & $-$2 \\
5 & dc204852\_148  & 14.44 0.04 & 13.74 0.06 & 13.44 0.04 & 0.70 0.07  & 0.30 0.07 & 1.00 0.06  & 0.0429 &   17.57 &  16.01  & 1 & $-$2 \\
6 & dc204852\_39   & 14.50 0.03 & 13.81 0.05 & 13.48 0.03 & 0.69 0.06  & 0.33 0.06 & 1.02 0.04  & 0.0489 &   17.77 &  16.16  & 1 & $-$2 \\
7 & dc204852\_45   & 15.04 0.04 & 14.39 0.05 & 14.03 0.03 & 0.65 0.06  & 0.36 0.06 & 1.01 0.05  & 0.0484 &   \nodata &  \nodata  & 1 & $-$2 \\
8 & dc204852\_104  & 14.99 0.05 & 14.25 0.05 & 13.95 0.04 & 0.74 0.07  & 0.30 0.06 & 1.04 0.06  & 0.0493 &   \nodata &  \nodata  & 1 &  0 \\
9 & dc204852\_149  & 13.91 0.03 & 13.24 0.05 & 12.91 0.04 & 0.67 0.06  & 0.33 0.06 & 1.00 0.05  & 0.0569 &   \nodata &  \nodata  & 1 &  0 \\
10 & dc204852\_192 & 13.83 0.05 & 13.13 0.04 & 12.80 0.05 & 0.70 0.06  & 0.33 0.06 & 1.03 0.07  & 0.0473 &   16.98 &  15.40  & 1 & $-$5 \\
11 & dc204852\_77  & 14.88 0.03 & 14.16 0.04 & 13.90 0.04 & 0.72 0.05  & 0.26 0.06 & 0.98 0.05  & 0.0452 &   \nodata &  \nodata  & 1 & $-$2 \\
12 & dc204852\_174 & 14.84 0.03 & 14.15 0.05 & 13.88 0.03 & 0.69 0.06  & 0.27 0.06 & 0.96 0.04  & 0.0448 &   18.09 &  16.43  & 1 & $-$5 \\
13 & dc204852\_184  & 14.29 0.02 & 13.60 0.04 & 13.25 0.04 & 0.69 0.04  & 0.35 0.06 & 1.04 0.04  & 0.0469 &   17.36 &  15.78  & 1 & $-$2 \\
14 & dc204852\_216  & 13.87 0.04 & 13.18 0.03 & 12.88 0.02 & 0.69 0.05  & 0.30 0.04 & 0.99 0.04  & 0.0490 &   \nodata &  \nodata  & 1 & $-$2 \\
15 & dc204852\_231  & 13.58 0.03 & 12.88 0.03 & 12.58 0.03 & 0.70 0.04  & 0.30 0.04 & 1.00 0.04  & 0.0459 &   16.72 &  15.18  & 1 & $-$2 \\
16 & dc032952\_135a & 14.34 0.02 & 13.52 0.04 & 13.09 0.04 & 0.82 0.04  & 0.43 0.06 & 1.25 0.04  & 0.0519 &   18.09 &  16.21  & 1 & $-$2 \\
17 & dc032952\_156a & 13.22 0.04 & 12.48 0.03 & 12.15 0.03 & 0.74 0.05  & 0.33 0.04 & 1.07 0.05  & 0.0604 &   16.61 &  14.93  & 1 & $-$2 \\
18 & dc010746\_30b  & 14.99 0.07 & 14.42 0.04 & 14.21 0.03 & 0.57 0.08  & 0.21 0.05 & 0.78 0.08  & 0.0267 &   17.90 &  16.41  & 1 & $-$5 \\
19 & dc032952\_82a  & 14.96 0.03 & 14.35 0.03 & 14.18 0.03 & 0.61 0.04  & 0.17 0.04 & 0.78 0.04  & 0.0576 &   17.81 &  16.42  & 1 & $-$5 \\
20 & dc032952\_158b & 14.13 0.03 & 13.41 0.02 & 13.02 0.04 & 0.72 0.04  & 0.39 0.04 & 1.11 0.05  & 0.0500 &   17.26 &  15.76  & 1 &  0 \\
21 & dc010746\_22m  & 14.49 0.04 & 13.87 0.04 & 13.60 0.02 & 0.62 0.06  & 0.27 0.04 & 0.89 0.04  & 0.0200 &   \nodata &  \nodata  & 1 &  0 \\
22 & dc010746\_45m  & 14.98 0.03 & 14.32 0.04 & 14.16 0.03 & 0.66 0.05  & 0.16 0.05 & 0.82 0.04  & 0.0300 &   17.37 &  16.24  & 1 &  0 \\
23 & ac103\_132     & 18.45 0.08 & 18.24 0.07 & 17.23 0.06 & 0.21 0.10  & 1.01 0.09 & 1.22 0.10  & 0.3047 &   \nodata &  19.34  & 2 &  6 \\
24 & ac114\_22      & 18.26 0.08 & 17.57 0.06 & 16.76 0.07 & 0.69 0.10  & 0.81 0.09 & 1.50 0.11  & 0.3354 &   \nodata &  19.85  & 2 &  0 \\
25 & ac114\_89      & 17.79 0.09 & 17.24 0.07 & NC    NC   & 0.55 0.11  & \nodata \nodata & \nodata \nodata & 0.3169 &   \nodata &  19.78  & 2 &  0 \\
26 & ac103\_03      & 16.33 0.08 & 15.44 0.08 & 15.09 0.05 & 0.89 0.11  & 0.35 0.09 & 1.24 0.09  & 0.3118 &   19.95 &  18.12  & 2 &  0 \\
27 & ac103\_106     & 17.15 0.09 & 16.34 0.07 & 15.76 0.06 & 0.81 0.11  & 0.58 0.09 & 1.39 0.11  & 0.3091 &   \nodata &  \nodata  & 2 &  0 \\
28 & ac103\_280     & 17.21 0.06 & 16.23 0.07 & 15.76 0.07 & 0.98 0.09  & 0.47 0.10 & 1.45 0.09  & 0.3111 &   21.00 &  18.93  & 2 &  0 \\
29 & ac103\_145     & 17.20 0.08 & 16.31 0.08 & 15.90 0.07 & 0.89 0.11  & 0.41 0.11 & 1.30 0.10  & 0.3105 &   \nodata &  19.66  & 2 &  3 \\
30 & lcrs01        & 16.18 0.04 & 15.57 0.05 & 15.10 0.04 & 0.61 0.06  & 0.47 0.06 & 1.08 0.06  & 0.0746 &   \nodata &  17.05  & 3 &  1 \\
31 & lcrs17        & 15.83 0.03 & 15.19 0.03 & 14.75 0.05 & 0.64 0.04  & 0.44 0.06 & 1.08 0.06  & 0.0609 &   \nodata &  16.99  & 3 &  0 \\
32 & lcrs21        & 15.55 0.03 & 14.94 0.04 & 14.53 0.04 & 0.61 0.05  & 0.41 0.06 & 1.02 0.05  & 0.0994 &   \nodata &  16.93  & 3 &  0 \\
33 & lcrs13        & 14.49 0.03 & 13.67 0.02 & 13.29 0.03 & 0.82 0.04  & 0.38 0.04 & 1.20 0.04  & 0.0957 &   \nodata &  12.97  & 3 &  1 \\
34 & lcrs14        & 14.90 0.03 & 14.20 0.05 & 13.77 0.03 & 0.70 0.06  & 0.43 0.06 & 1.13 0.04  & 0.0704 &   \nodata &  16.05  & 3 &  0 \\
35 & lcrs12        & 15.02 0.03 & 14.35 0.04 & 13.82 0.03 & 0.67 0.05  & 0.53 0.05 & 1.20 0.04  & 0.0971 &   \nodata &  16.78  & 3 &  1 \\
36 & lcrs03        & 14.11 0.04 & 13.47 0.03 & 13.04 0.03 & 0.64 0.05  & 0.43 0.04 & 1.07 0.05  & 0.0810 &   \nodata &  15.03  & 3 &  1 \\
37 & lcrs16        & 15.35 0.04 & 14.75 0.04 & 14.41 0.03 & 0.60 0.06  & 0.34 0.05 & 0.94 0.05  & 0.0764 &   \nodata &  16.69  & 3 &  1 \\
38 & lcrs15        & 15.84 0.05 & 15.16 0.05 & 14.73 0.04 & 0.68 0.07  & 0.43 0.06 & 1.11 0.06  & 0.1137 &   \nodata &  17.19  & 3 &  0 \\
39 & lcrs06        & 15.64 0.05 & 15.09 0.04 & 14.72 0.03 & 0.55 0.06  & 0.37 0.05 & 0.92 0.06  & 0.0884 &   \nodata &  16.81  & 3 &  0 \\
40 & lcrs08        & 15.63 0.04 & 15.01 0.03 & 14.55 0.04 & 0.62 0.05  & 0.46 0.05 & 1.08 0.06  & 0.1121 &   \nodata &  17.87  & 3 & $-$2 \\
41 & lcrs07        & 13.62 0.05 & 12.89 0.03 & 12.45 0.04 & 0.73 0.06  & 0.44 0.05 & 1.17 0.06  & 0.1141 &   \nodata &  15.00  & 3 &  0 \\
42 & lcrs20        & 14.48 0.03 & 13.89 0.03 & 13.53 0.05 & 0.59 0.04  & 0.36 0.06 & 0.95 0.06  & 0.0632 &   \nodata &  15.96  & 3 & $-$2 \\
43 & lcrs18        & 14.70 0.04 & 14.02 0.03 & 13.62 0.03 & 0.68 0.05  & 0.40 0.04 & 1.08 0.05  & 0.0598 &   \nodata &  16.09  & 3 &  0 \\
44 & lcrs05        & 15.36 0.05 & 14.80 0.03 & 14.32 0.05 & 0.56 0.06  & 0.48 0.06 & 1.04 0.07  & 0.1172 &   \nodata &  16.73  & 3 & $-$2 \\
45 & lcrs19        & 14.95 0.03 & 14.24 0.04 & 13.90 0.03 & 0.71 0.05  & 0.34 0.05 & 1.05 0.04  & 0.0640 &   \nodata &  16.42  & 3 &  0 \\
46 & lcrs11        & 15.48 0.04 & 14.78 0.04 & 14.38 0.03 & 0.70 0.06  & 0.40 0.05 & 1.10 0.05  & 0.1216 &   \nodata &  16.96  & 3 &  0 \\
47 & lcrs02        & 14.95 0.03 & 14.28 0.03 & 13.95 0.03 & 0.67 0.04  & 0.33 0.04 & 1.00 0.04  & 0.0987 &   \nodata &  16.36  & 3 &  2 \\
48 & lcrs09        & 15.98 0.05 & 15.30 0.04 & 14.96 0.03 & 0.68 0.06  & 0.34 0.05 & 1.02 0.06  & 0.0651 &   \nodata &  17.47  & 3 &  0 \\
49 & lcrs10        & 15.29 0.04 & 14.65 0.05 & 14.27 0.04 & 0.64 0.06  & 0.38 0.06 & 1.02 0.05  & 0.1049 &   \nodata &  16.68  & 3 &  0 \\
50 & lcrs04        & 14.49 0.04 & 13.80 0.05 & 13.41 0.03 & 0.69 0.06  & 0.39 0.06 & 1.08 0.05  & 0.1012 &   \nodata &  15.68  & 3 &  1 \\
51 & pgc35435      & 11.75 0.03 & 10.98 0.04 & 10.66 0.02 & 0.77 0.05  & 0.32 0.04 & 1.09 0.04  & 0.0178 &   13.75 &  \nodata  & 4 & $-$3 \\
52 & dc204852\_116 & 12.62 0.06 & 11.92 0.05 & 11.65 0.02 & 0.70 0.08  & 0.27 0.05 & 0.97 0.06  & 0.0441 &  15.84 &  14.06  & 4 & $-$5 \\
53 & dc204852\_66  & 14.45 0.05 & 13.62 0.04 & 13.36 0.03 & 0.83 0.06  & 0.26 0.05 & 1.09 0.06  & 0.0410 &  17.48 &  15.88  & 4 & $-$5 \\
54 & pgc60102      & 12.96 0.06 & 12.13 0.03 & 11.65 0.04 & 0.84 0.07  & 0.47 0.05 & 1.31 0.07  & 0.0304 &   15.36 &  \nodata  & 4 & $-$2 \\
55 & eso290-IG\_050& 13.46 0.03 & 12.74 0.03 & 12.39 0.03 & 0.72 0.04  & 0.35 0.04 & 1.07 0.04  & 0.0290 &  15.18 &  14.21  & 4 & $-$2 \\
56 & pgc62615      & 12.65 0.04 & 11.92 0.04 & 11.63 0.04 & 0.73 0.06  & 0.29 0.06 & 1.02 0.06  & 0.0280 &   \nodata &  \nodata  & 4 &  2 \\
57 & pgc57612      & 10.99 0.03 & 10.22 0.03 & 10.10 0.03 & 0.77 0.04  & 0.11 0.04 & 0.88 0.04  & 0.0183 &   13.30 &  11.33  & 4 & $-$5 \\
58 & ngc6653       & 11.53 0.04 & 10.79 0.01 & 10.59 0.04 & 0.74 0.04  & 0.20 0.04 & 0.94 0.06  & 0.0172 &   \nodata &  \nodata  & 4 & $-$5 \\
59 & dc204852\_115 & 14.98 0.03 & 14.33 0.03 & 14.04 0.03 & 0.65 0.04  & 0.29 0.04 & 0.94 0.04  & 0.0440 &  18.13 &  16.53  & 4 & $-$5 \\
60 & dc204852\_126 & 15.01 0.04 & 14.29 0.04 & 14.01 0.04 & 0.72 0.06  & 0.28 0.06 & 1.00 0.06  & 0.0489 &  18.21 &  16.60  & 4 & $-$2 \\
61 & dc204852\_38  & 13.49 0.05 & 12.90 0.03 & 12.56 0.04 & 0.59 0.06  & 0.34 0.05 & 0.93 0.06  & 0.0454 &  16.73 &  15.12  & 4 & $-$2 \\ 
62 & ngc6328       & 11.33 0.03 & 10.57 0.04 & 10.24 0.04 & 0.77 0.05  & 0.32 0.06 & 1.09 0.05  & 0.0142 &  13.17 &  11.45  & 4 &  2 \\ 
63 & pgc62765      & 11.42 0.04 & 10.68 0.04 & 10.36 0.03 & 0.74 0.06  & 0.32 0.05 & 1.06 0.05  & 0.0193 &   \nodata &  \nodata  & 4 & $-$2 \\
\tablenotetext{(1)}{Correlative number.}
\tablenotetext{(2)}{Name of the galaxy.}
\tablenotetext{(3)}{$J_s$ apparent magnitude and photometric error.}
\tablenotetext{(4)}{$H$ apparent magnitude and photometric error.}
\tablenotetext{(5)}{$K_s$ apparent magnitude and photometric error.}
\tablenotetext{(6)}{$J - H$ color index and its error.}
\tablenotetext{(7)}{$H - K_s$ color index and its error.}
\tablenotetext{(8)}{$J - K_s$ color index and its error.}
\tablenotetext{(9)}{Redshift.}
\tablenotetext{(10)}{$B$ apparent total magnitude in the Johnson system. This magnitude
	is provided by NED.}
\tablenotetext{(11)}{$R$ apparent total magnitude in the Cousins system. This
	magnitude is provided by NED.}
\tablenotetext{(12)}{Sample ID (see Table \ref{samples}).} 
\tablenotetext{(13)}{Morphological T-type provided by NED.}
\enddata
\end{deluxetable}
\clearpage

\figcaption[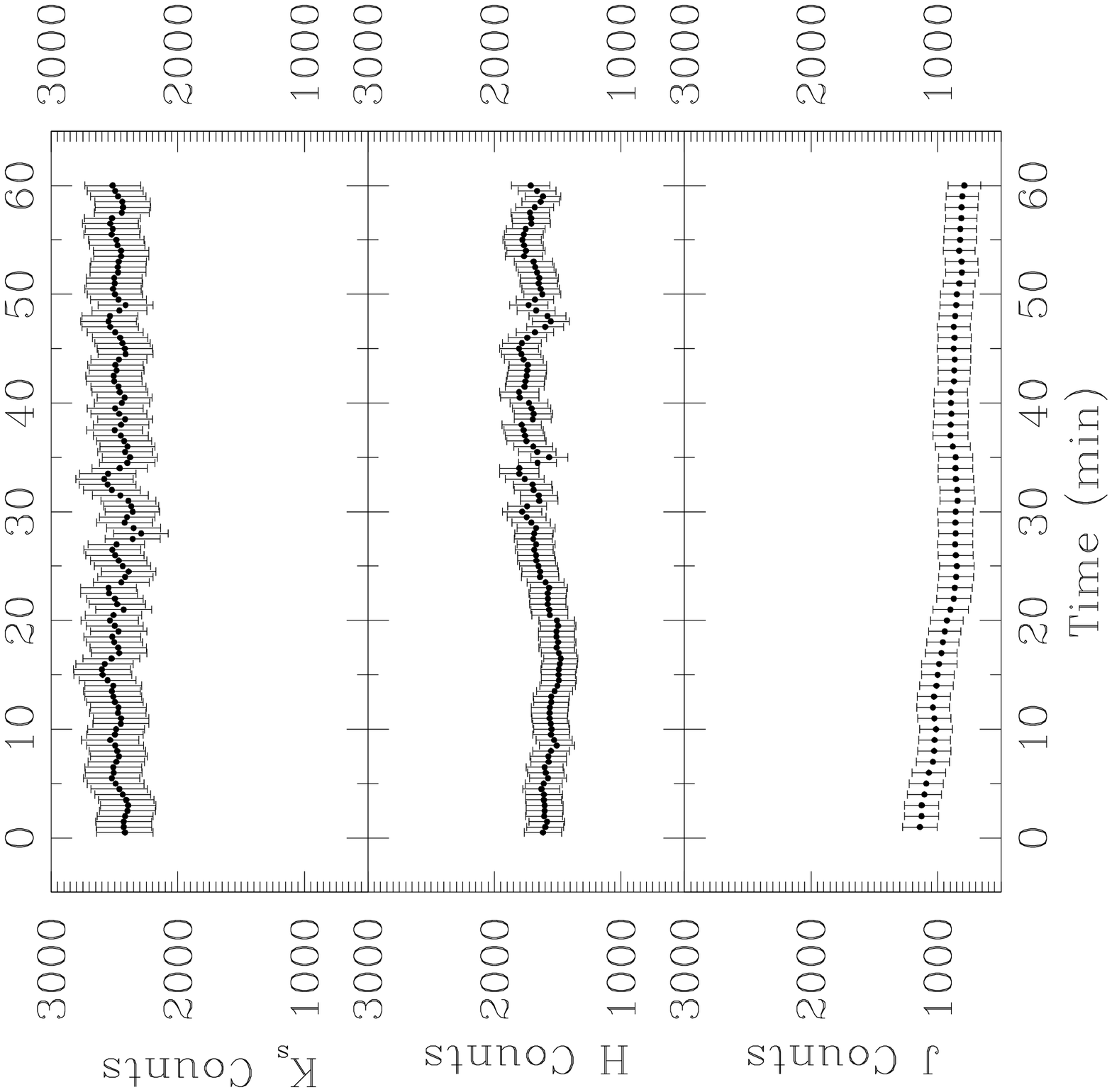]{Typical sky 
variations in the near-IR passbands 
$J$, $H$, and $K_s$ during 1 hour. Each point represents the mean sky
level for an individual image during an {\em observing sequence}. The error bars
are given by the standard deviation in the image counts. Note the 
larger error bars for the $K_s$ band, where the thermal variations are
in fact larger (where also the shot noise is higher, compared to that
of other filters). This behavior limits the accuracy of the sky subtraction 
procedure applied to the near-IR images (see text). \label{sky}}

\figcaption[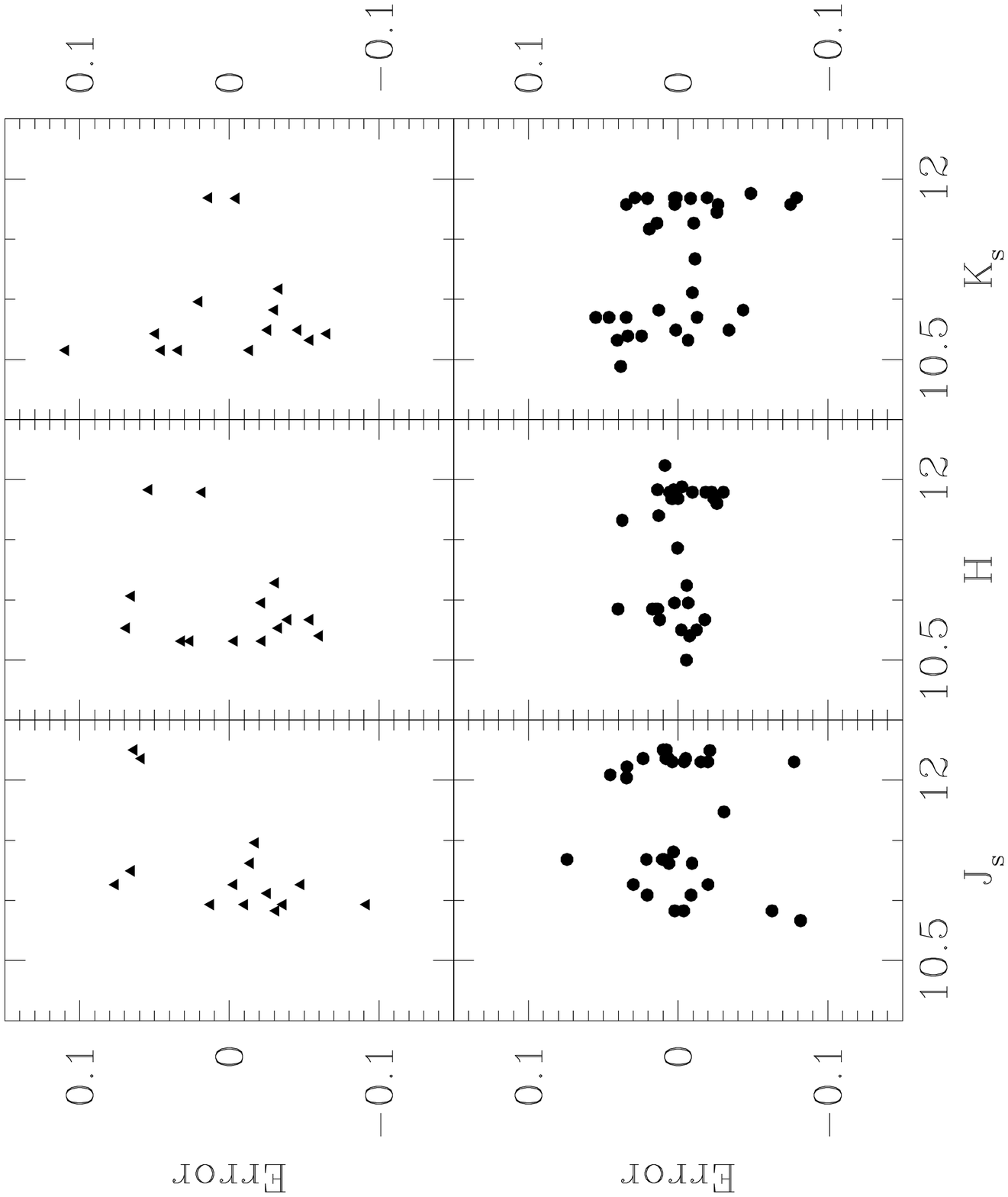]{Average errors 
in the photometric calibrations for the standards
observed with the 100-inch du Pont telescope (triangles), and for the 
standards observed with the 40-inch Swope telescope (circles). 
Every point corresponds to an average error of several (typically no less than 3)
measurements for the same standard, observed in different nights.
\label{stds_error}}

\figcaption[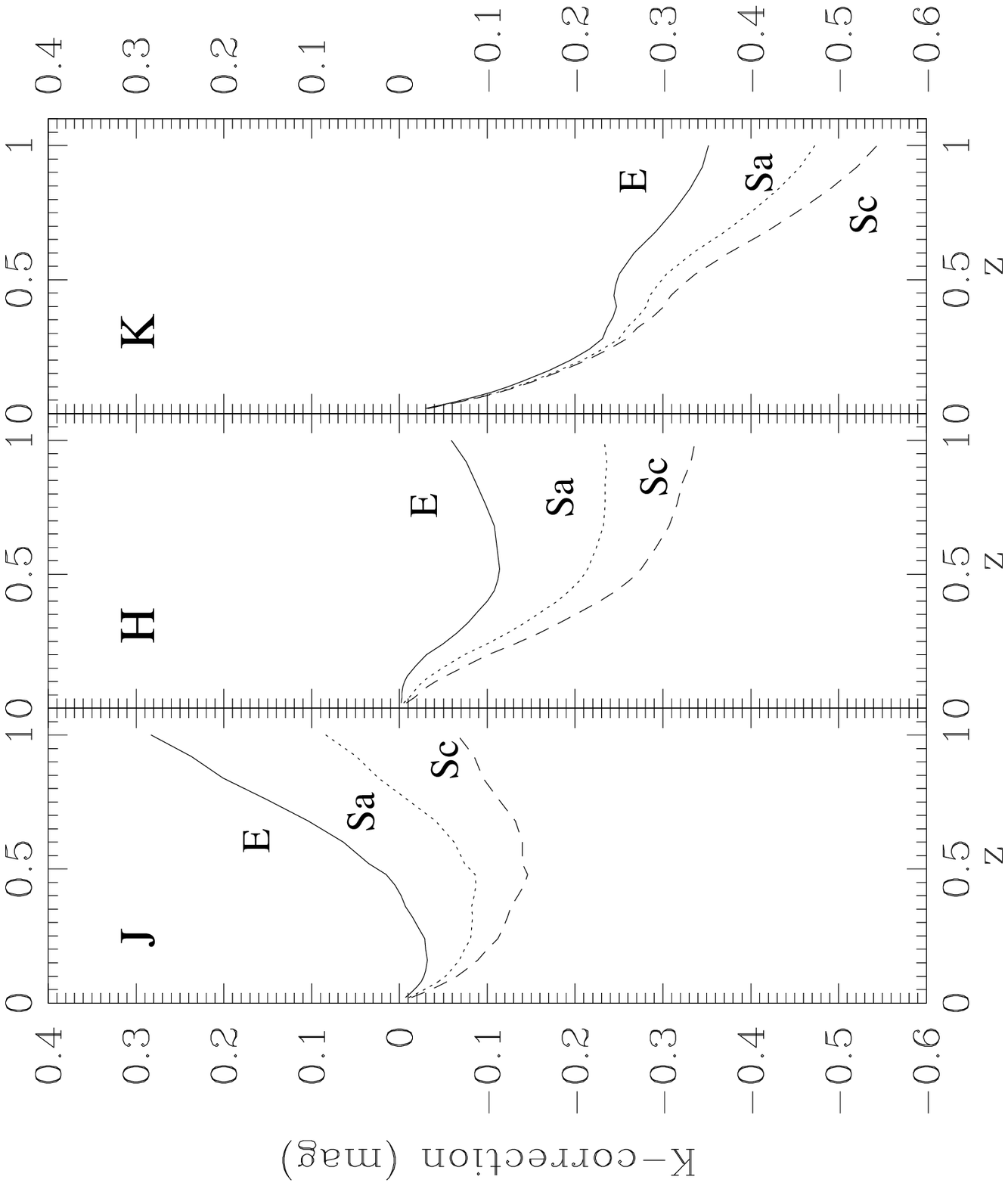]{K-corrections 
from the models of \citet{poggianti97} 
for passbands $J$, $H$ and $K$, as
a function of redshift and for Hubble types E, Sa and Sc.
Note the large and negative K-corrections for
the $K$ band for $z \gtrsim 0.5$. \label{K_poggianti}}

\figcaption[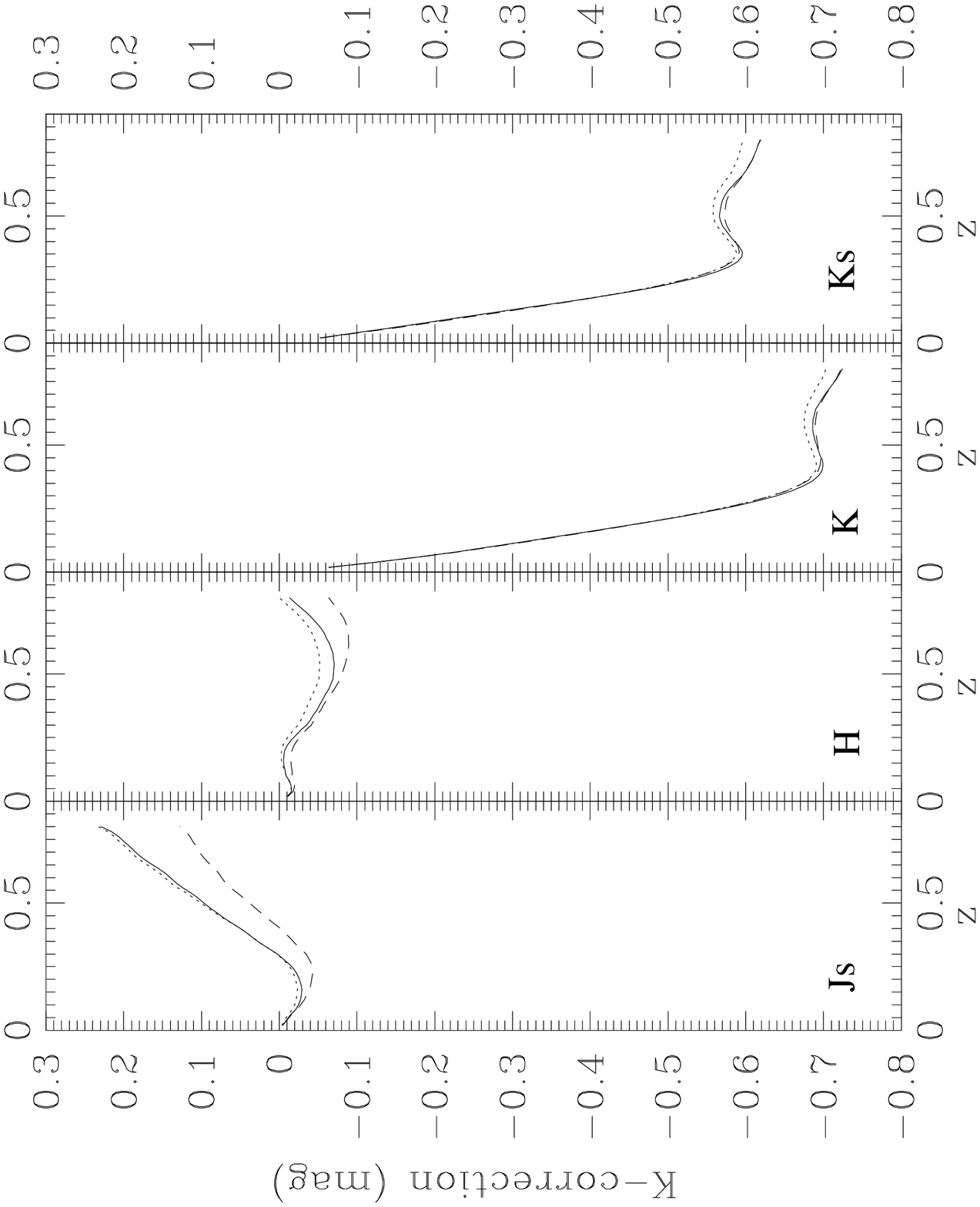]{K-corrections 
derived from the PEGASE models \citep{fioc97} 
for the $J$, $H$, $K$ and $K_s$ bands as 
a function of redshift and Hubble type (lines as in Figure 
\ref{K_poggianti}). Compare with Figure \ref{K_poggianti}. \label{K_pegase}}

\figcaption[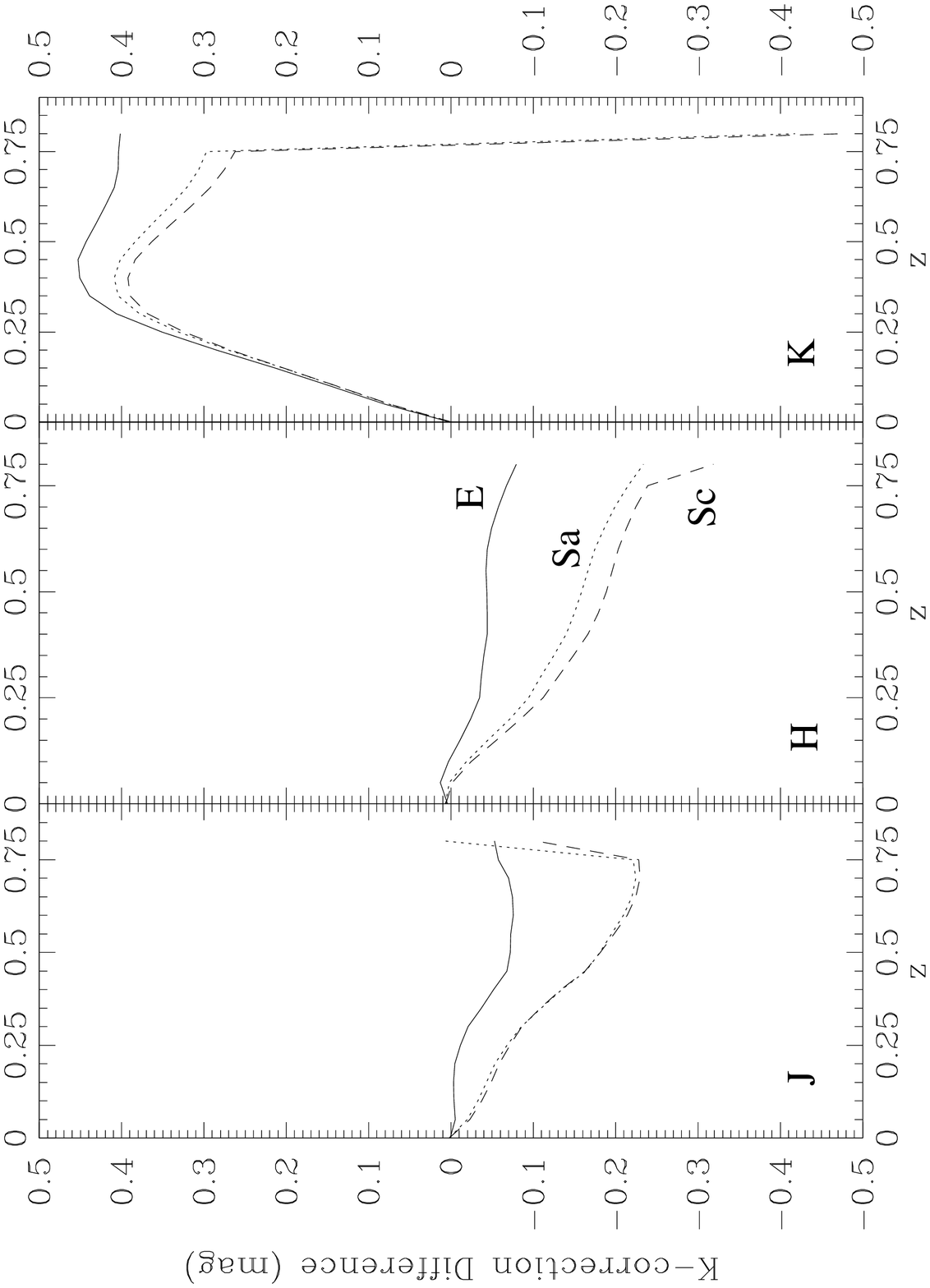]{K-correction
differences obtained from 2 spectrophotometric models of galaxy evolution, 
indicated in Figures \ref{K_poggianti} and \ref{K_pegase}. Differences are computed
for $J$, $H$, and $K$ and for Hubble types E, Sa and Sc. Note the large
difference for the K-corrections in the $K$ band. \label{K_diff}}

\figcaption[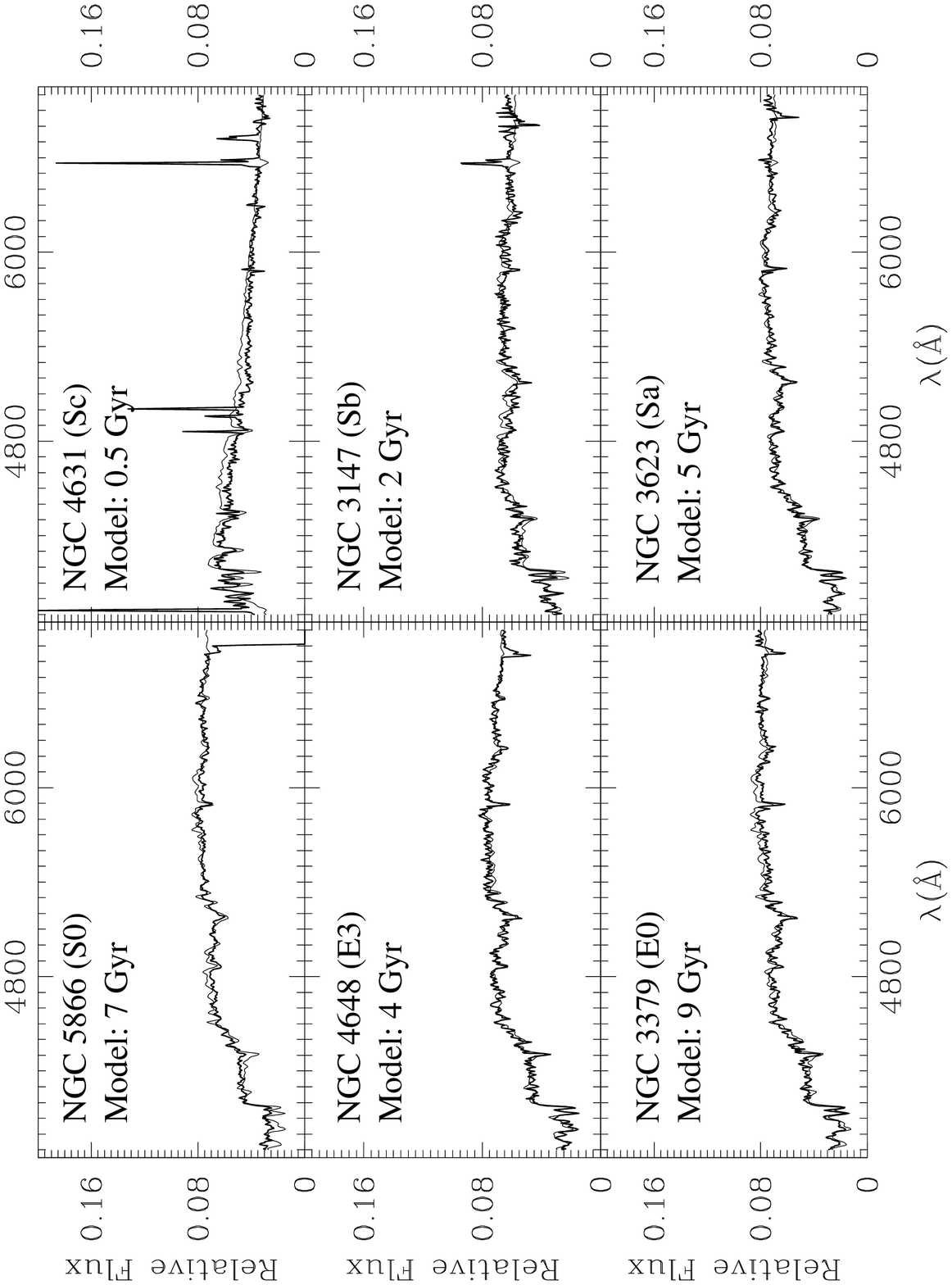]{Some \citet{kennicutt92} spectra
of observed normal galaxies (as indicated in each panel) 
with known Hubble types (thick lines), and fitted synthetic
spectra from GISSEL96 \citep{charlot96} (thin lines). The fitted models correspond
to instantaneous bursts of solar metallicity and different ages of the passively
evolving stellar populations. The closest model spectrum 
is obtained using a simple $\chi^2$ fitting algorithm between the Kennicutt spectra and 20 
model spectra. The good match shows that at the 
optical wavelengths models agree with observations. The same kind of models are
compared to the near-IR colors. See text for details. \label{spectra}}

\figcaption[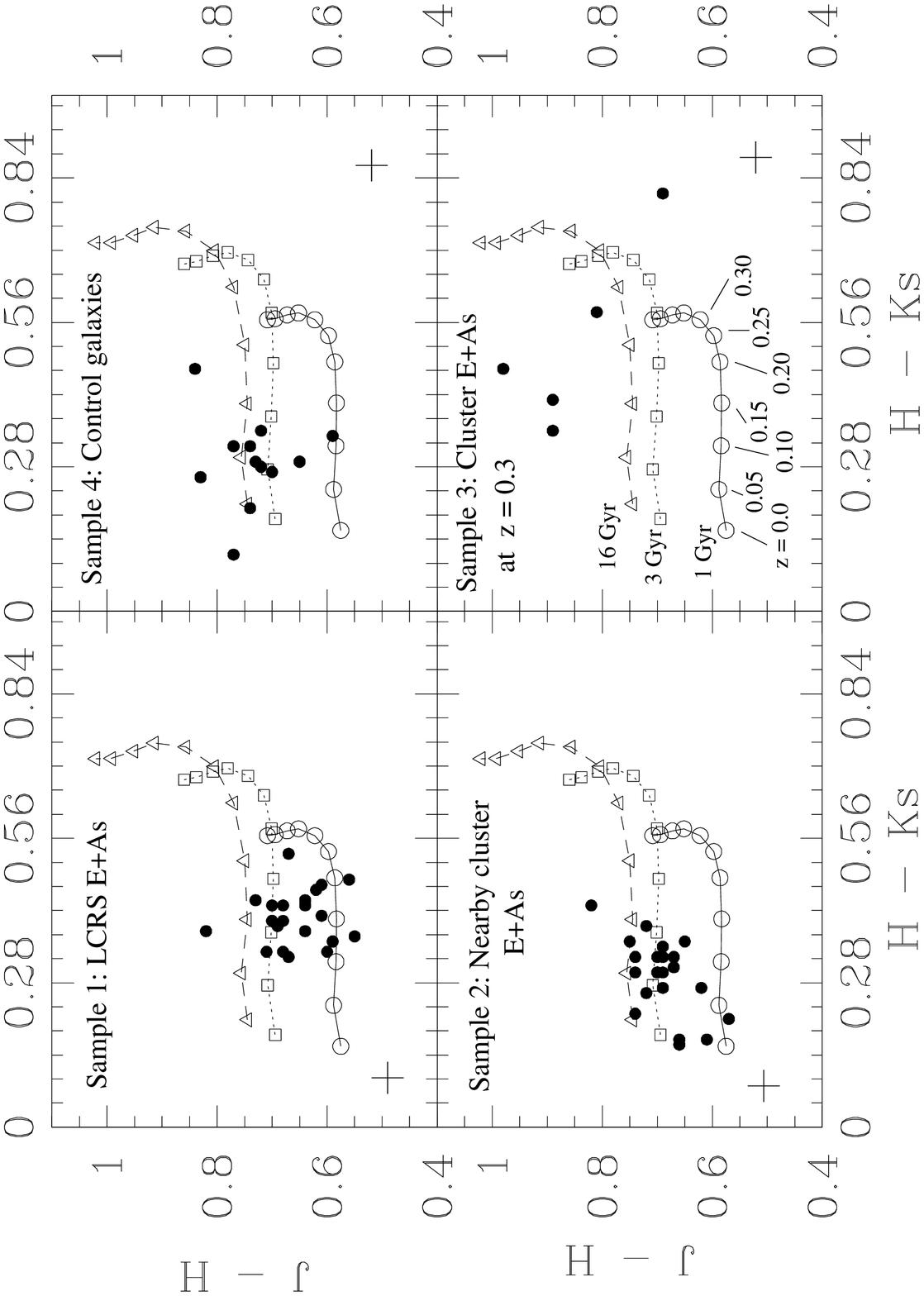]{Observed colors of the 
E+A galaxies reported in this paper (filled circles) compared with spectrophotometric models
of galaxy evolution (open symbols joint by lines). Each panel corresponds to a different
E+A sample (as indicated in each panel). Each line represent a redshift track of 
an instantaneous burst of solar metallicity, at a given age of 1, 3 and 16 Gyr, 
indicated by circles, squares, and triangles, respectively, for the 
redshifts indicated in the lower right panel. The crosses are the error bars in the
colors. See text for explanations \label{obs_colors}}

\figcaption[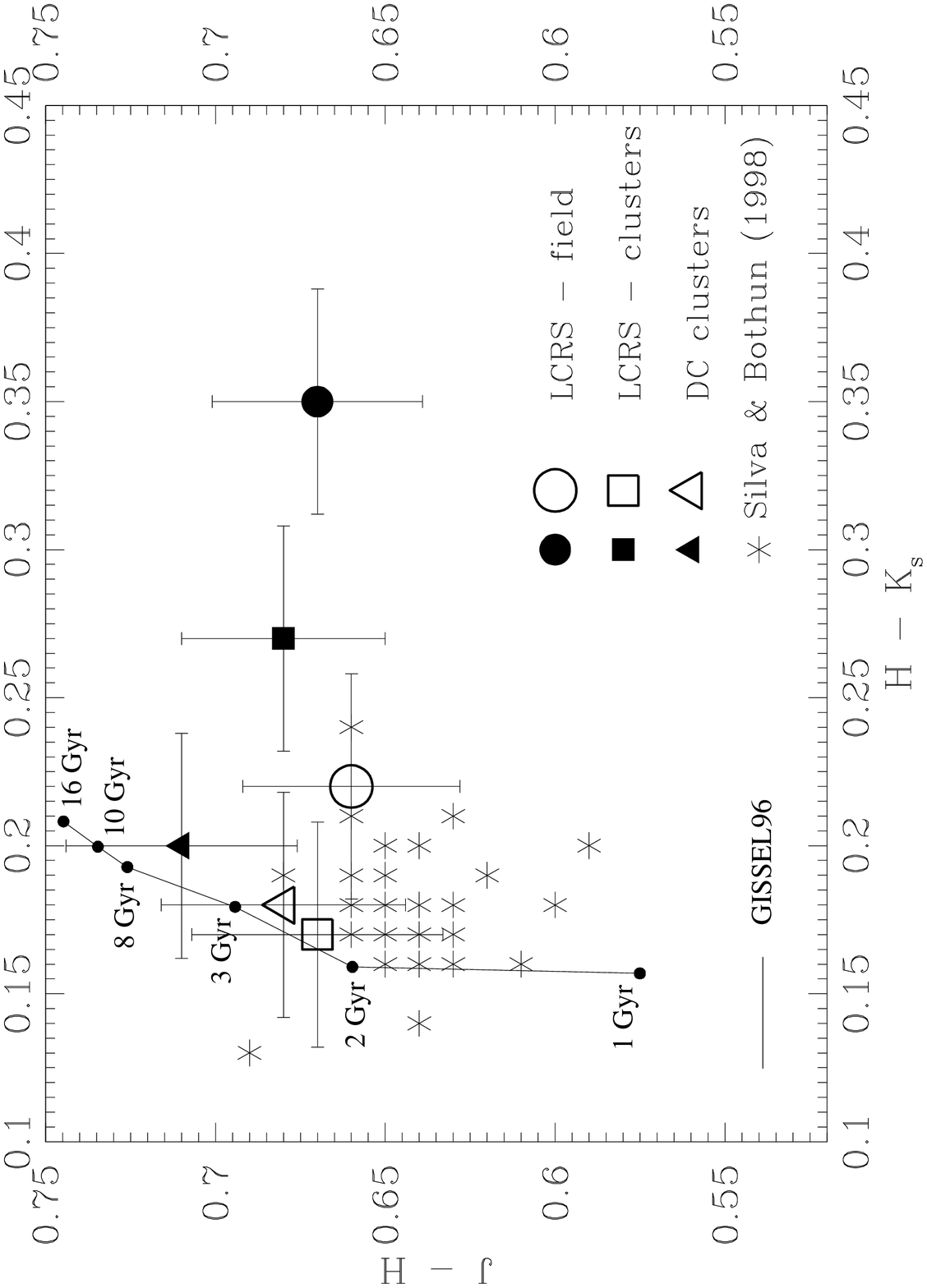]{Averaged rest-frame 
colors of E+As lying in different environments. The LCRS symbols 
corresponds to the 21 E+As from the 
sample of Las Campanas Redshift Survey \citep{zabludoff96}. Most of these galaxies 
are located in the field (at $<z> \sim 0.1$), but 3 of them lie in clusters. 
The DC cluster E+As correspond to the E+As from the sample of 
\citet{caldwell97}, and all of them are located in clusters with $<z> \sim 0.05$. 
Filled symbols indicate that rest-frame colors have been obtained using the 
\citet{poggianti97} K-corrections. Open symbols are averaged rest-frame colors
obtained using the PEGASE \citep{fioc97} K-corrections. Asteriscs correspond 
to elliptical galaxies observed by \citet{silva98}. The solid line 
correspond to colors of a GISSEL96 \citep{charlot96} instantaneous burst of
solar metallicity at $z = 0$ and at different ages (indicated by solid dots and
labeled). See text for details. \label{color_average}}

\figcaption[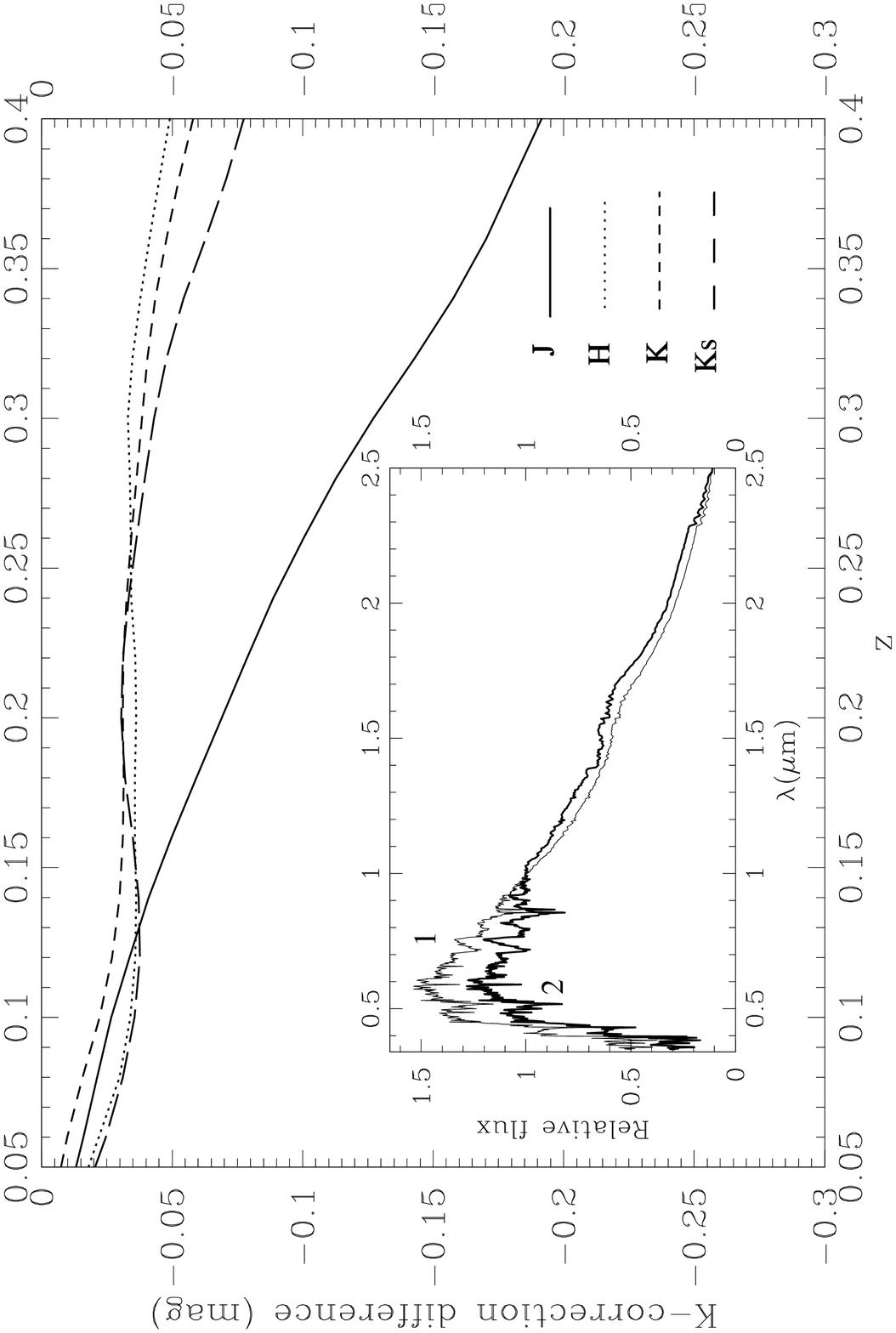]{K-correction differences in $J$, $H$,
$K$ and $K_s$, as a function of redshift, for two SEDs having different metallicity. 
The two SEDs, shown in the inset, are simple instantaneous bursts with a Scalo 
initial mass function \citep{scalo86} and with an age of 10 Gyr. Differences in K-corrections
are expressed as the difference between K-correction for SED 1 ([Fe/H] $= -0.30$) 
and K-correction for SED 2 ([Fe/H] $= +0.10$). \label{diff_metal}}

\begin{figure}
\plotone{galaz.fig1.ps}
\end{figure}
\begin{figure}
\plotone{galaz.fig2.ps}
\end{figure}
\begin{figure}
\plotone{galaz.fig3.ps}
\end{figure}
\begin{figure}
\plotone{galaz.fig4.ps}
\end{figure}
\begin{figure}
\plotone{galaz.fig5.ps}
\end{figure}
\begin{figure}
\plotone{galaz.fig6.ps}
\end{figure}
\begin{figure}
\plotone{galaz.fig7.ps}
\end{figure}
\begin{figure}
\plotone{galaz.fig8.ps}
\end{figure}
\begin{figure}
\plotone{galaz.fig9.ps}
\end{figure}

\end{document}